\documentclass[twocolumn,showpacs,amsmath,amssymb,bibnotes,floatfix]{revtex4}
\usepackage{bm}
\usepackage{graphicx}
\usepackage{epsfig}
\usepackage{subfigure}
\usepackage{afterpage}
\usepackage{bm}
\usepackage{amsfonts}
\usepackage{color}

\newcommand{\be}{\begin{equation}}
\newcommand{\ee}{\end{equation}}
\newcommand{\bea}{\begin{eqnarray}}
\newcommand{\eea}{\end{eqnarray}}
\newcommand{\gdot}{\dot{\gamma}}
\newcommand{\gdotbar}{\overline{\dot{\gamma}}}

\newcommand{\bw}{\begin{widetext}}
\newcommand{\ew}{\end{widetext}}

\begin{document}

\title{Nonlinear dynamics and rheology of active fluids: simulations in two dimensions} \author{S. M. Fielding,$^1$ D. Marenduzzo,$^2$
  M. E. Cates$^2$} \affiliation{ $^1$Department of Physics, Durham University, Science
  Laboratories, South Road, Durham DH1 3LE, UK\\ $^2$SUPA, School of Physics and Astronomy,
  University of Edinburgh, JCMB Kings Buildings, Edinburgh EH9 3JZ,
  UK
 }

\date{\today}
\begin{abstract}

{We report simulations of a continuum model for (apolar, flow aligning) active 
fluids in two dimensions. Both free and anchored boundary conditions are 
considered, at parallel confining walls that are either static or moving at 
fixed relative velocity. We focus on extensile materials and find that steady 
shear bands, previously shown to arise ubiquitously in 1D for the active 
nematic phase at small (or indeed zero) shear rate, are generally replaced in 
2D by more complex flow patterns that can be stationary, oscillatory, or 
apparently chaotic. The consequences of these flow patterns for time-averaged 
steady-state rheology are examined. }

\end{abstract}
      
\maketitle


\section{Introduction}
\label{sec:intro}

Active systems in general, and active fluids in particular, have recently 
become a topical research area in soft matter physics~\cite{Ramaswamy,Toner,kruse1,nedelec,Voituriez,Aranson,goldstein,Marenduzzo07,Cates08,Giomi08,Giomi10,Baskaran09,EPL,Liverpool,Liverpool06,ramaswamy3,llopis,ignacio,SoftMatterReview}. 
An ``active particle'' is a particle which consumes energy from the
surrouding environment to do work. In many cases this work is used for 
self-propulsion. An active fluid is then a suspension of such
active particles in an underlying Newtonian fluid. The distinguishing feature 
of activity in this context is the ability of the particle to exert forces on 
the surrounding fluid. In the absence of body forces applied externally, the 
simplest perturbation which a particle can impart on the fluid is a force 
dipole. According to the direction of the force pair making up the dipole, a
single particle can be either ``extensile'' -- if the forces are exerted
from the centre of mass to the fluid -- or ``contractile'' -- if it is
exerted from the fluid to the centre of mass of the particle. (For a rodlike 
particle, extensile forcing ejects fluid in both directions along the axis and 
draws it in around the equator; the reverse is true for contractile motion.)
Also, the dipole can be applied at the centre of drag of the object or 
elsewhere -- only in the latter case can this active forcing lead to
motility (creating a ``mover'' as opposed to a ``shaker'' \cite{Ramaswamy}). 
The presence of a force dipole in active fluids defines a director at the 
particle level even in cases where the organism is not rodlike. Thus one can
introduce a natural order parameter which characterises
the magnitude of local orientational order in the fluid. As a result, in 
concentrated active fluids one generically expects an isotropic-nematic 
transition, and this is why such systems are commonly described by means of 
equations of motion that closely resemble those of liquid crystalline 
materials. 

A suspension of bacterial swimmers, such as {\it B. subtilis}, is
an example of an extensile active fluid, whereas a suspension of algae like {\it Chlamydomonas} is contractile. At a quite different length scale, the same equations can be used to describe an actomyosin solution containing filaments and motors. In this case the active motion of the motors, which resemble moving cross-links on the network of filaments, creates a contractile effect. Such actomyosin networks represent a simplified picture of the cytoskeletal structures underlying the motility of eukaroytic cells \cite{nedelec,kruse1}. In this work we prefer the term
``active fluids'' to ``active gels'' for all the systems under study, although the latter term is widely used. (In fact not all these materials
have a markedly viscoelastic rheological response, as the term ``gel'' suggests; we shall discuss this later on.)  The very definition of active particles implies that they are far from equilibrium; unsurprisingly therefore, nonequilibrium phenomena are a dominant theme in the physics of active fluids.
 
For instance, it has been predicted theoretically~\cite{Voituriez}
that active fluids should undergo a nonequilibrium phase transition
between a ``passive'' quiescent phase, where the motion of each of the 
particles are basically uncorrelated and the coarse grained mean velocity
field is uniform and zero, as in conventional passive unforced fluids, and
an ``active'' phase, in which long-range correlations lead to a 
non-zero ``spontaneous flow'' in steady state. This transition has also
been seen experimentally: there the spontaneous flow patterns have been
mapped out by following the dynamics of individual bacteria in a 
thin {\it B. subtilis} film, where the local concentration of bacteria
was very high~\cite{Aranson}. What triggers
the transition between the quiescent and the ``active'' phase in these
bacterial fluids is simply the increase in concentration -- which controls, among other things, the magnitude of  an ``activity'' parameter which appears in the equations of motion that we will
consider later on. The bacterial flow patterns invariably were seen to
involve vortices and swirls of bacteria, similar to those which 
are observed in aerobic bacteria which move around to get into contact with
oxygen~\cite{goldstein}. These large-scale motions are sometimes referred to
as ``bacterial turbulence'' because the flow field resembles that 
of a Newtonian fluid at high Reynolds number, in the turbulent regime.
However, it should be kept in mind that, unlike the turbulent flow
of Newtonian fluids, that of active fluids occurs at essentially 
zero Reynolds number, so that inertia plays no role.

Although less characterised experimentally as yet, the rheology of active 
fluids is also expected to display a very intriguing 
phenomenology~\cite{Liverpool06,Marenduzzo07,Cates08,Giomi08}.
Extensile and contractile fluids lead to very different rheological 
responses. It was predicted and later on confirmed by simulations in 1D that
contractile active gels should show an almost solid-like behaviour when
sheared, and that if left free to reorient at the boundary they should
possess a non-zero yield stress~\cite{Liverpool06,Cates08}.
Extensile fluids on the other hand should have in 1D a ``negative yield stress'' causing a window of apparently superfluid rheology (in which a nonzero macroscopic shear rate can be accommodated at zero stress)
close to the transition between the isotropic and the nematic phase.

An outstanding problem in the {continuum} theory of active fluids is that most of
the calculations done so far assume a 1D (or rather a quasi-1D) geometry, in which the
variations of the orientational order parameter and the flow field
are limited to just one dimension. (The orientational order itself lives in a 2D or 3D space.) While this is enough to 
describe the spontaneous flow transition and to map out rheology curves,
it is legitimate to ask how much the 1D predictions are confirmed by
fully 2D or 3D simulations. 
(The same would hold in most fluid dynamics problems involving the occurrence of
hydrodynamic instabilities; spontaneous flow is just one of these.)
A small number of simulations in 2D for
extensile spontaneously flowing fluids, within an active but isotropic phase, have already shown that flow
patterns can indeed be significantly different in 2D than in 1D. For instance, while
in 1D a spontaneous net flow is the first state found on entering the
active phase, in 2D rolls and turbulent flow were instead observed, which are
more closely related to the experiments in Ref.~\cite{goldstein,Aranson}.

{Although, as stated above, there are few works in 2D simulating active fluids using the hydrodynamic continuum equations of \cite{Ramaswamy,kruse1} as we pursue here, several other simulation avenues have been pursued in 2D. These include a two-phase model, developed by Wolgemuth \cite{Wolgemuth} in a context specific to bacteria; this focuses on the coupling between activity and number density, which we neglect in this paper. Other contributions have involved simulation of discrete swimmers in two or three dimensions \cite{Ishikawa,Santillan,Graham,llopis,Rupert}; however, such approaches become increasingly difficult as one approaches the case of dense swarms for which the hydrodynamic description (without coupling to number density) was primarily developed. Several of these methods have shown onset of bacterial turbulence or development of coherent structures of various kinds, but it is not yet clear, for instance, which of these require the coupling of activity to number density.} 

Our aim in this paper is therefore to study in more detail {the simplest continuum models for} 2D flowing
states of active fluids, focusing on the extensile case, 
close to or within the nematic phase, both in the absence of
any forcing and when subjected to a small shear rate. Besides, we evaluate
the role of boundary conditions by comparing simulations in which the
nematic order parameter field is fixed on the surface (to ensure planar
alignment of the associated director field) to others in which the
director is free to rotate at the boundary planes. We confirm that
2D simulations of spontaneous flow patterns in the active phase 
markedly differ from 1D patterns. We also show that the rheology curves
may differ when the 1D approximation is lifted. In essence, this paper 
extends to 2D the work presented for 1D systems in our earlier paper with 
Orlandini and Yeomans \cite{Cates08}. Naturally one can ask what would happen if one generalizes further to the fully 3D case. A comprehensive study of this case would be computationally exhausting, and we defer it to future work. However, some initial progress has recently been made using stability analyses to guide a selective exploration of parameter space \cite{Morozov}. 

The rest of this paper is structured as follows. In Sec.~\ref{sec:equations}
we review the equations of motion of apolar active fluids, to which we
restrict ourselves here. In Sec.~\ref{sec:geometry} we specify for our simulations the
geometry and boundary conditions; the
parameter values and units used; and (briefly) the algorithmic methods involved. In
Sec.~\ref{sec:1D} we summarise and extend slightly our earlier results~\cite{Cates08}
for the 1D hydrodynamics and rheology of these active suspensions.
The results of our new 2D simulation study are then presented in
Sec.~\ref{sec:extensile} -- note that we focus on active extensile
materials which lead to spontaneous flow in a quasi-1D geometry,
for rod-like active particles~\cite{Cates08,Voituriez}. 
Sec.~\ref{sec:conclusion} contains
conclusions and perspectives for future study.

\section{Model equations}
\label{sec:equations}

Following standard procedures \cite{SoftMatterReview} we first employ a Landau -- de Gennes free energy ${\cal F} = \int fdV$ to describe an equilibrium isotropic to nematic transition in a material without activity.  The free energy density can be written as
a sum of two terms, $f=f_1+f_2$.  The first is a bulk contribution
\begin{eqnarray}
{f}_1=\frac{A_0}{2}(1 - \frac {\gamma} {3}) Q_{\alpha \beta}^2 -
          \frac {A_0 \gamma}{3} Q_{\alpha \beta}Q_{\beta
          \gamma}Q_{\gamma \alpha}
+ \frac {A_0 \gamma}{4} (Q_{\alpha \beta}^2)^2
\label{eqBulkFree}
\end{eqnarray}
while the second is a distortion term, which we take in a (standard) one-constant approximation as \cite{degennes}
\begin{equation}\label{distorsion}
{f}_2=\frac{K}{2} \left(\partial_\gamma Q_{\alpha \beta}\right)^2.
\end{equation}
In the equations above, $A_0$ is a constant measuring the relative 
contributions of the bulk and the distortion term, $\gamma$ controls the
magnitude of order (it may be viewed as an effective temperature or
concentration for thermotropic and lyotropic liquid crystals
respectively), while $K$ is an elastic constant.  The resulting free energy density $f$ is standard to describe passive nematic liquid
crystals \cite{degennes}.  Here and in what follows Greek indices
denote cartesian components and summation over repeated indices is
implied.

The equation of motion for {\bf Q} is taken to be \cite{beris,O92,O99}
\begin{equation}
(\partial_t+{\bf u}\cdot{\bf \nabla}){\bf Q}-{\bf S}({\bf \nabla u},{\bf
    Q})= \Gamma {\bf H}+\lambda {\bf Q}
\label{Qevolution}
\end{equation}
where $\Gamma$ is a collective rotational diffusion constant. We have added a term proportional to 
$\lambda$, which is one of two ``activity parameters'' that become nonzero when activity is switched on. However, this term is easily absorbed into a redefinition of the coefficient $\gamma$ in Eq. (\ref{eqBulkFree}) and from now on we suppress it: $\lambda = 0$.
{The form of Eq. (\ref{Qevolution}) was suggested on the basis of
  symmetry in Refs. \cite{Ramaswamy,kruse1} and derived starting from
  an underlying microscopic model in Ref. \cite{Liverpool}.}  The
first term on the left-hand side of Eq. (\ref{Qevolution}) is the
material derivative describing the usual time dependence of a quantity
advected by a fluid with velocity ${\bf u}$. This is generalized for
rod-like molecules by a second term
\begin{eqnarray}\label{S_definition}
{\bf S}({\bf \nabla u},{\bf Q})
& = &(\xi{\bf D}+{\bf \omega})({\bf Q}+{\bf I}/3)
\\ \nonumber 
& + & ({\bf Q}+
{\bf I}/3)(\xi{\bf D}-{\bf \omega}) \\ \nonumber
& - & 2\xi({\bf Q}+{\bf I}/3){\mbox{Tr}}({\bf Q}{\bf \nabla u})
\end{eqnarray}
where ${\mbox{Tr}}$ denotes the tensorial trace, while ${\bf D}=({\bf \nabla u}+{\bf
  \nabla u}^T)/2$ and ${\bf \omega}=({\bf \nabla u}-{\bf \nabla u}^T)/2$ are the symmetric
part and the anti-symmetric part respectively of the velocity gradient
tensor $\nabla u_{\alpha\beta}=\partial_\beta u_\alpha$.  The constant $\xi$
depends on the molecular details of a given particle and controls whether the passive material is flow-aligning or flow-tumbling. To isolate the nolinear effects of activity we choose flow-aligning materials in this paper (bearing in mind that flow-tumbling ones often show unsteady flow behaviour even without activity).  The first
term on the right-hand side of Eq. (\ref{Qevolution}) describes the
relaxation of the order parameter towards the minimum of the free
energy. The molecular field ${\bf H}$ which provides the force for
this motion is given by
\begin{equation}
{\bf H}= -{\delta {\cal F} \over \delta {\bf Q}}+({\bf
    I}/3) \mbox{Tr}{\delta {\cal F} \over \delta {\bf Q}}.
\label{molecularfield}
\end{equation}

The fluid velocity, ${\bf u}$, obeys the continuity equation for an effectively incompressible fluid, $\nabla.{\bf u}= 0$, and also
the corresponding Navier-Stokes equation,
\begin{equation}\label{navierstokes}
\rho(\partial_t+ u_\beta \partial_\beta)
u_\alpha = \partial_\beta (\Pi_{\alpha\beta})+
\eta \partial_\beta(\partial_\alpha
u_\beta + \partial_\beta u_\alpha)
\end{equation}
Here $\rho$ is the fluid density, $\eta$ is a viscosity and
$\Pi_{\alpha\beta}=\Pi^{\rm passive}_{\alpha\beta}+ \Pi^{\rm
  active}_{\alpha\beta}$. {Note that this viscosity $\eta$ in Eq.\ref{navierstokes} is isotropic, and that of the solvent in which our active particles are suspended. This does not preclude the emergence of an anisotropic macroscopic viscosity as a result of the coupling to those particles.}
The stress tensor
$\Pi^{\rm passive}_{\alpha\beta}$ necessary to describe ordinary LC
hydrodynamics without activity is:
\begin{eqnarray}\label{BEstress}
\Pi^{\rm passive}_{\alpha\beta}= &-&P_0 \delta_{\alpha \beta} +2\xi
(Q_{\alpha\beta}+{1\over 3}\delta_{\alpha\beta})Q_{\gamma\epsilon}
H_{\gamma\epsilon}\\\nonumber &-&\xi
H_{\alpha\gamma}(Q_{\gamma\beta}+{1\over 3}\delta_{\gamma\beta})-\xi
(Q_{\alpha\gamma}+{1\over
  3}\delta_{\alpha\gamma})H_{\gamma\beta}\\ \nonumber
&-&\partial_\alpha Q_{\gamma\nu} {\delta {\cal F}\over
  \delta\partial_\beta Q_{\gamma\nu}} +Q_{\alpha \gamma} H_{\gamma
  \beta} -H_{\alpha \gamma}Q_{\gamma \beta}. 
\end{eqnarray}
%
The nematic free energy gives an effectively constant contribution to the fluid pressure $P_0$ (whose final spatial variation is determined by the fluid incompressibility condition).  The active stress, which unlike the passive one cannot be derived from any underlying free energy functional, is given to leading order by
\begin{equation}
\Pi^{\rm active}_{\alpha\beta}=-\zeta  Q_{\alpha\beta} \label{activestress}
\end{equation}
where $\zeta$ is a second activity constant \cite{Ramaswamy,EPL}.
Note that with the sign convention chosen here $\zeta>0$ corresponds
to extensile rods and $\zeta<0$ to contractile ones \cite{Ramaswamy}; in either case this term does not merely renormalize the equations for a passive liquid crystal but fundamentally alters their form. Accordingly it is a key control parameter in the continuum description of active nematics.
As in Eq. \ref{Qevolution}, the explicit form of the active contribution to the stress tensor entering Eq. \ref{navierstokes} was proposed on the basis of a symmetry analysis of a fluid of contractile or extensile dipolar objects in \cite{Ramaswamy}. It was also derived by coarse graining a more microscopic model for a solution of actin fibers and myosins in Ref. \cite{Liverpool}.

The equations of motion chosen above address the case of active nematics, by 
which we mean particles whose local ordering is apolar. This means that 
locally one has a preferred orientation of the force dipole but not of any 
vector field such as the mean swimming direction of motile particles. 
The equations of motion in the latter case are yet more complicated 
(see e.g. \cite{SoftMatterReview}) but because the active stress takes the 
same form as above, for rheological purposes are expected to yield broadly 
similar results. In this work, for simplicity, we study solely the apolar case.
 

\section{Simulation details}
{
In this Section we describe the geometry and boundary conditions used for our numerical work and also discuss the units and parameter values chosen, and the issue of numerical convergence. Note that (primarily for historical reasons) we used finite difference (FD) for free boundaries and lattice Boltzmann (LB) for fixed ones; this is considered further in Appendix A.}

\subsection{Geometry and boundary conditions}
\label{sec:geometry}

We consider a slab of active fluid sandwiched between flat parallel
plates located at $y=0,L_y$. The fluid velocity and order parameter tensor can vary in both $x$ and $y$, while in the third
dimension $z$ we assume translational invariance. 
Each plate has length $L_x$ and periodic boundary conditions are imposed in the $x$ direction. In general the
upper plate is taken to move in the positive $x$ direction at a
constant speed $\gdotbar L_y$, although many of the results presented below are for the case with no externally applied shear,
$\gdotbar=0$. At the plates we impose boundary conditions of no slip and no permeation for the fluid velocity. For the molecular order parameter, we will
study two different boundary conditions:

\begin{itemize}

\item So-called ``free'' boundary conditions, in which the order parameter tensor at the wall can take any value but must satisfy a zero-gradient condition in the direction normal to the wall:
\be
\partial_y Q_{\alpha\beta}=0,\;\;\; {\rm at}\;\;\; y=\{0,L_y\}\;\;\;\forall\;{\alpha,\beta}\label{eqn:free}
\ee
\end{itemize}
Free boundaries are believed to give the fastest convergence to bulk behavior by minimizing the effects of the confining walls on the order parameter dynamics.

\begin{itemize}
\item So-called ``fixed'' boundary conditions, in which {the 
director field is anchored along the $x$ direction parallel to the wall}, and the order parameter tensor $Q_{\alpha\beta}$ has the equilibrium form for a uniaxial state with this director:
\be 
Q_{\alpha\beta} = q\left(n_{\alpha}n_{\beta}-\frac{\delta_{\alpha\beta}}{3}
\right)
\ee
where
\be
q= \frac{1}{4} + \frac{3}{4}\sqrt{1- \frac{8}{3\gamma}}
\ee
is the magnitude of order in the surface, which we take to be equal to 
the one in the bulk. 

\end{itemize}

The choice of fixed boundary conditions is motivated by the behavior of 
non-active liquid crystals which frequently have specific anchoring 
interactions that lock in the director field at the boundary. 
Their appropriateness for active nematics is less well established 
(particularly for bacterial suspensions, although some sort of anchoring 
remains plausible for actomysin gels), which is why a comparison of the two 
types of boundary condition is appropriate here. In fact, as we will see 
below, our main conclusions concerning the flow patterns and rheology are 
relatively robust to the choice of boundary conditions (although these 
certainly influence some of the details).

\subsection{Units and parameter values}
\label{sec:parameters}

Throughout our study we use units of length in which the gap between
the plates $L_y=1$; units of time in which the model's underlying microscopic timescale $\tau\equiv
1/(A_0\Gamma)=1$; and units of mass such that the free energy parameter $A_0$, which has the dimensions of a stress, likewise obeys $A_0 =1$. 
All runs use a value of the I/N control parameter
$\gamma=3.0$ which lies within the nematic phase for a system without 
activity; as stated previously this value effectively absorbs the activity 
parameter $\lambda$, which we have set to zero. Throughout we set $\xi=0.7$, 
which corresponds to a typical value for flow-aligning molecular nematics {\cite{degennes}; such nematics generally comprise rod-like molecules of modest aspect ratio. In any case}   
we expect that, so long as one does not leave the flow-aligning regime, our 
results should be relatively insensitive to this choice. In the system of 
units just defined, we choose for the Newtonian viscosity a value 
$\eta=0.567$, which is much larger than the viscoelastic one without activity,
hence leading to a conventional Newtonian rheology in the passive phase. 

With the above choices, and denoting by $\gdotbar$ the mean shear rate imposed 
by the relative motion of the two plates, the parameter values that remain to 
be varied between runs are the activity level $\zeta$, and a microscopic
lengthscale  on which elastic distortions compete 
with bulk free energies. This obeys $l\equiv \sqrt{K/A_0} = \sqrt{K}$ (where 
the latter equality holds in our system of units only). Accordingly, we 
will quote values for the parameters $(\zeta,l)$ in each figure caption
of our results sections below. A third parameter, the Reynolds number (Re) 
which controls the relative strength of inertial to viscous terms in the 
Navier-Stokes equation, is always small or zero (see below).

{
\subsection{Numerical convergence}
\label{sec:convergence}
}
In a class of systems whose flow frequently is (or appears to be) chaotic, one cannot expect genuine convergence of numerically calculated velocity patterns with respect to the
time-step $\Delta t$ and mesh scale $(\Delta y, \Delta x)$.
We strive to ensure our results are converged, in the sense that further refinement gives no significant change to the type of behavior observed (except very close to parameter values at which there is transition from one regime to another), nor to macroscopic quantities like the time-averaged stress. In the units chosen above, we found this to typically require values of $(\Delta y, \Delta x, \Delta t)=(1/100,\,1/100,\,0.34)$
in our LB simulations and $(\Delta y, \Delta x, \Delta
t)=(1/128,\,4/512,\,0.05)$ -- or sometimes $(\Delta y, \Delta x, \Delta
t)=(1/256,\,4/1024,\,0.05)$ -- in our FD simulations. For sufficiently small $l$, one expects the boundary conditions to be unimportant so that (modulo the minor technical differences discussed above) the results from FD and LB should approach one another in this limit.

\section{Results in zero and one dimension}
\label{sec:1D}

\begin{figure}[tbp]
  \includegraphics[scale=0.6]{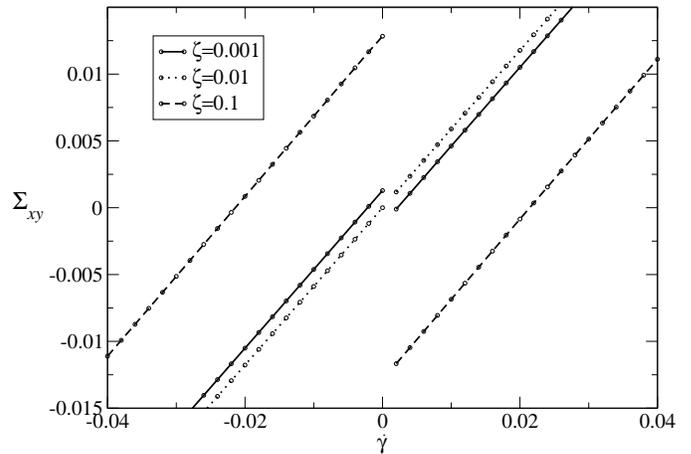}
  \caption{Homogeneous constitutive curves.}
\label{fig:fc}
\end{figure}

\begin{figure}[tb]
  \includegraphics*[scale=0.6]{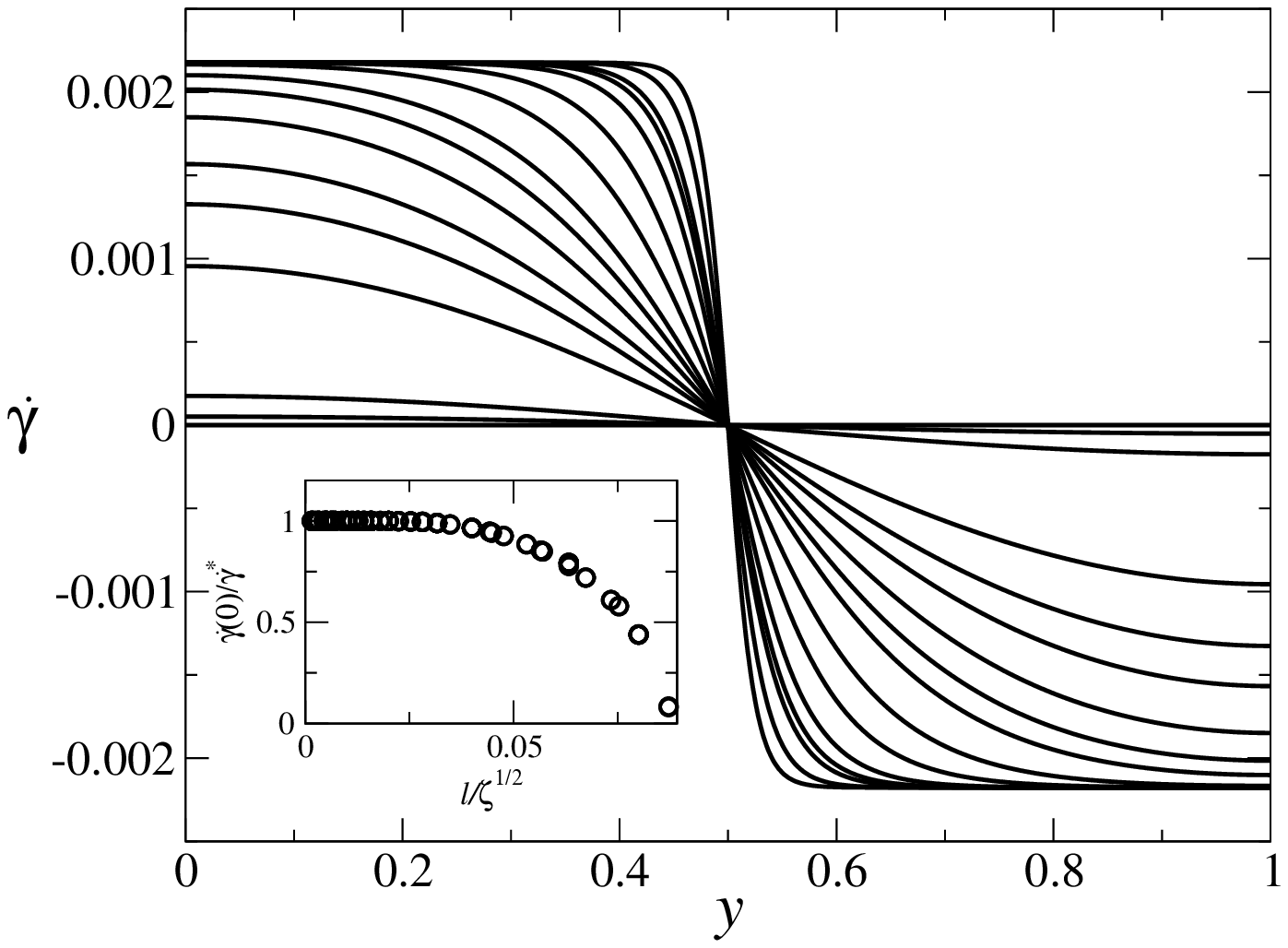}
  \includegraphics*[scale=0.6]{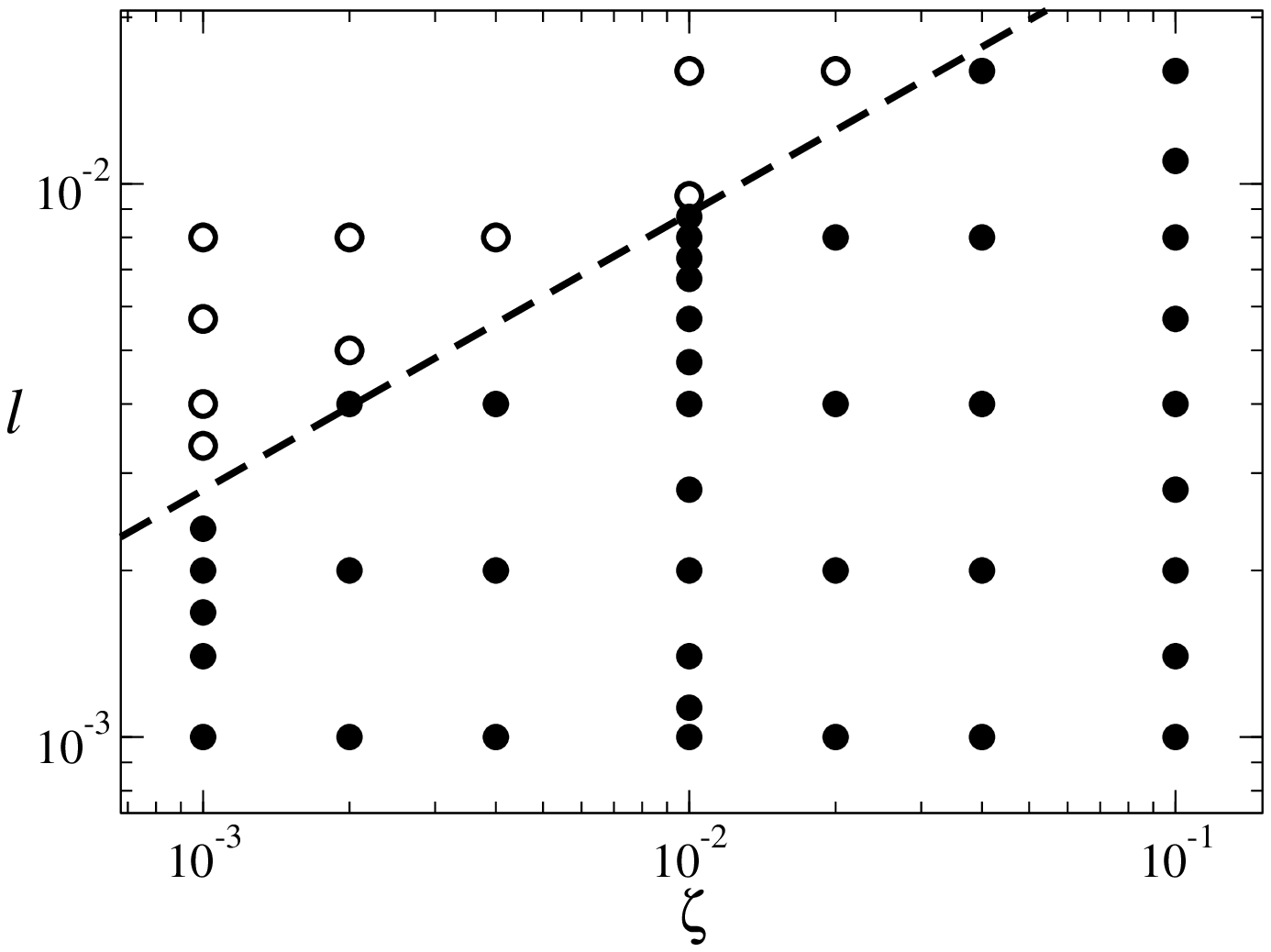}
  \caption{{\bf Top}: shear banded (or quiescent) profiles for
    $\zeta=0.01$. Values of $l=$0.0005, 0.00071, 0.001, 0.00113,
    0.0014, 0.002, 0.0028, 0.004, 0.00476, 0.0057, 0.00673, 0.0073,
    0.008, 0.00872, 0.00951, 0.016 increasing for decreasing
    $\gdot(y=0)$. Inset: shear rate $\gdot(y=0)$ at the left edge of
    the cell extracted from such profiles, for various values of
    $\zeta,l$, shown in the master scaling representation discussed in
    the main text. \newline {\bf Bottom:} Phase diagram for 1D runs
    showing the region of spontaneous shear banded flow (closed
    circles) and the region in which an initially heterogeneous state
    decays back to a quiescent state of zero flow (open circles). The
    dashed line is the power law $l^* = a \zeta^{1/2}$ using the value
    of the intercept $a$ extracted from the master scaling plot in the
    inset to the top figure.}
\label{fig:profiles1D}
\end{figure}


In this section we briefly recap some results of Ref.~\cite{Cates08} for the
hydrodynamics and rheology of an extensile fluid in less than two dimensions. 
We discuss first the case of a homogeneous (0D) imposed
shear flow, and then recall the results of calculations that allow spatial
variations in one spatial dimension (1D), $y$, assuming translational
invariance in both $z$ and $x$.

The homogeneous constitutive curves are shown in Fig.~\ref{fig:fc} for
three different values of positive (extensile) activity, $\zeta > 0$.
As discussed in Ref.~\cite{Cates08}, the vertical drop at the origin  arises 
from the alignment by weak flow of the nematic director at the Leslie angle 
(this occurs, for a passive system, throughout the nematic phase). This 
alignment produces an active stress tensor whose $xy$ component for extensile 
particles is of opposite sign with respect to shear rate.
This negative stress contribution is proportional to 
$\zeta$ and depends on the sign of the imposed flow $\dot\gamma$ but not its 
magnitude (when this is small). This discontinuous variation creates a 
downward step in the flow curve, which is expected
to allow the spontaneous formation of shear bands, even when no
external shear is applied at the system's boundaries, 
$\gdotbar=0$~\cite{Cates08}.

Accordingly, we perform a series of unsheared 1D runs in which spatial
variations are allowed in $y$ (only). Each is initialised with the
tensor $Q_{\alpha\beta}(y)$ having a uniform ($y$-independent) degree
of uniaxial order, and its long axis in the $xy$ plane at an angle
$\theta=22^\circ \tanh( (0.5-y)/l)$ to the $x$ direction. This favours
the formation of just two shear bands. (For random initial conditions,
multiple bands can in principle form in this planar flow geometry.)
The code was then evolved to steady state.

In the limit $l\to 0$ for any value of $\zeta$ we find the steady
state to comprise two coexisting bands of equal and opposite shear
rates $\pm\gdot^*(\zeta)$, these being the values at which the two
positively sloping branches of the homogeneous constitutive curve
intercept the $\gdot$ axis (see Fig.~\ref{fig:profiles1D}).  The bands
are separated from each other by a slightly diffuse interface of
thickness $l$. Increasing $l$ in a series of runs at any fixed value
of $\zeta$ eventually eliminates this state of spontaneous flow, with
quiescence being restored at a critical value $l^*=a\zeta^{1/2}$ where
$a\approx 0.885$. This is seen in the master scaling curve of
$\gdot(y=0,t\to\infty)/\gdot^*$ versus $l/\zeta^{1/2}$ in the inset of
Fig.~\ref{fig:profiles1D}, top, and by the dashed line separating
quiescence (open circles) from spontaneous banded states (filled
circles) in Fig.~\ref{fig:profiles1D}, bottom.

The results just presented were found with free boundary conditions. 
Simulations of the fixed boundary case give a more complex behavior, 
particularly at the larger values of $l$, where the anchoring condition at the 
wall can compete with the elastic director distortions required to maintain the 
shear-banded state. For a fuller discussion of such effects in 1D 
see \cite{Cates08}.

\section{2D Results: extensile systems}
\label{sec:extensile}

The remainder of the paper concerns 2D simulations that allow spatial
variations parallel to the plates (along $x$) as well as perpendicular
to them (along $y$). Furthermore we focus on extensile
fluids, which lead to spontaneous flow in 1D (contractile systems 
will be treated elsewhere).

\subsection{Unsheared systems}
\label{sec:unsheared}

We discuss first systems that are not subject to any externally
applied shear flow, treating the free and fixed boundary condition cases in 
turn. (Systems with applied shear will be addressed in 
Sec.~\ref{sec:sheared} below.) 

\subsubsection{Free boundary conditions}

\begin{figure}[tb]
  \includegraphics[scale=0.53]{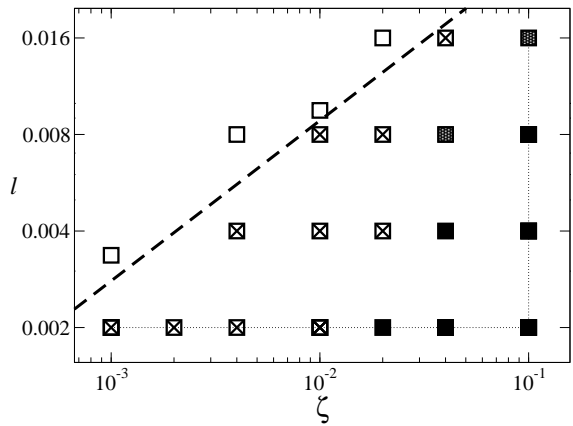}
  \caption{Phase diagram for 2D runs with free boundary conditions, each denoted by a square. Empty
    squares: quiescence. Elsewhere initial shear banded profile gives
    way to a state of 2D domains that is: steady (crosses);
    oscillatory (shaded square); aperiodic (filled squares).}
\label{fig:phaseDiagram2D}
\end{figure}

\if{
\begin{figure}[tb]
  \includegraphics[scale=0.53]{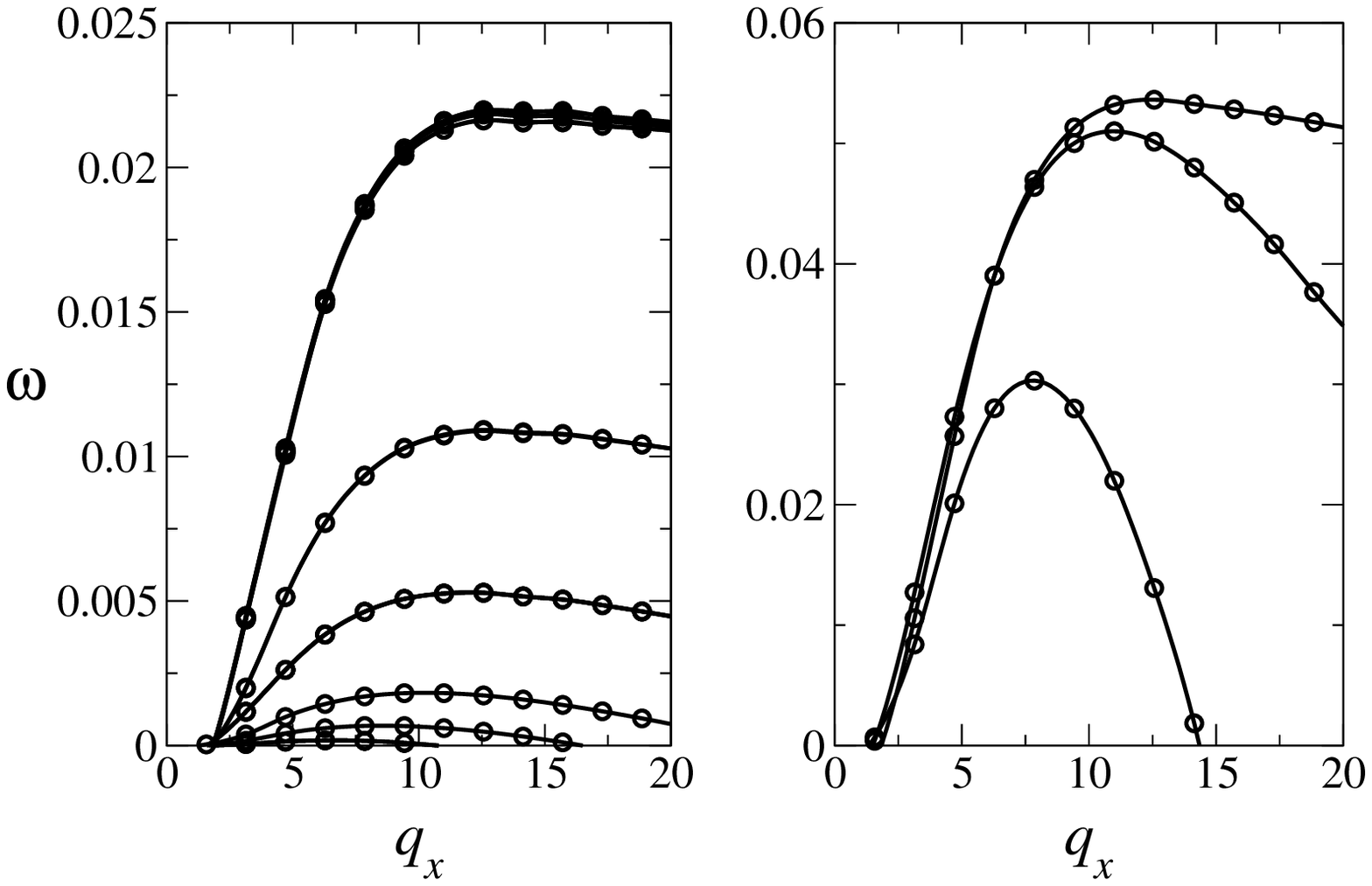}
  \caption{Dispersion relation for initial growth rate of 2D
    perturbations as a function of wavevector $q_x$. Left hand graph
    shows $l=0.002$ and $\zeta=0.001,0.002,0.004,0.01,0.02,0.04$
    (curve sets upwards). Three curves are shown for
    $\zeta=0.001,0.01,0.04$ (default, half mesh size, half
    time-step). For $\zeta=0.001,0.01$ they are indistinguishable,
    showing that convergence has been attained. For $\zeta=0.04$ they are
    almost indistinguishable.  Right hand graph: $\zeta=0.1$ and
    $l=0.004,0.008,0.016$ (curves
    downwards).}
\label{fig:dispersion}
\end{figure}
}\fi

\begin{figure}[tb]
  \includegraphics[scale=0.325]{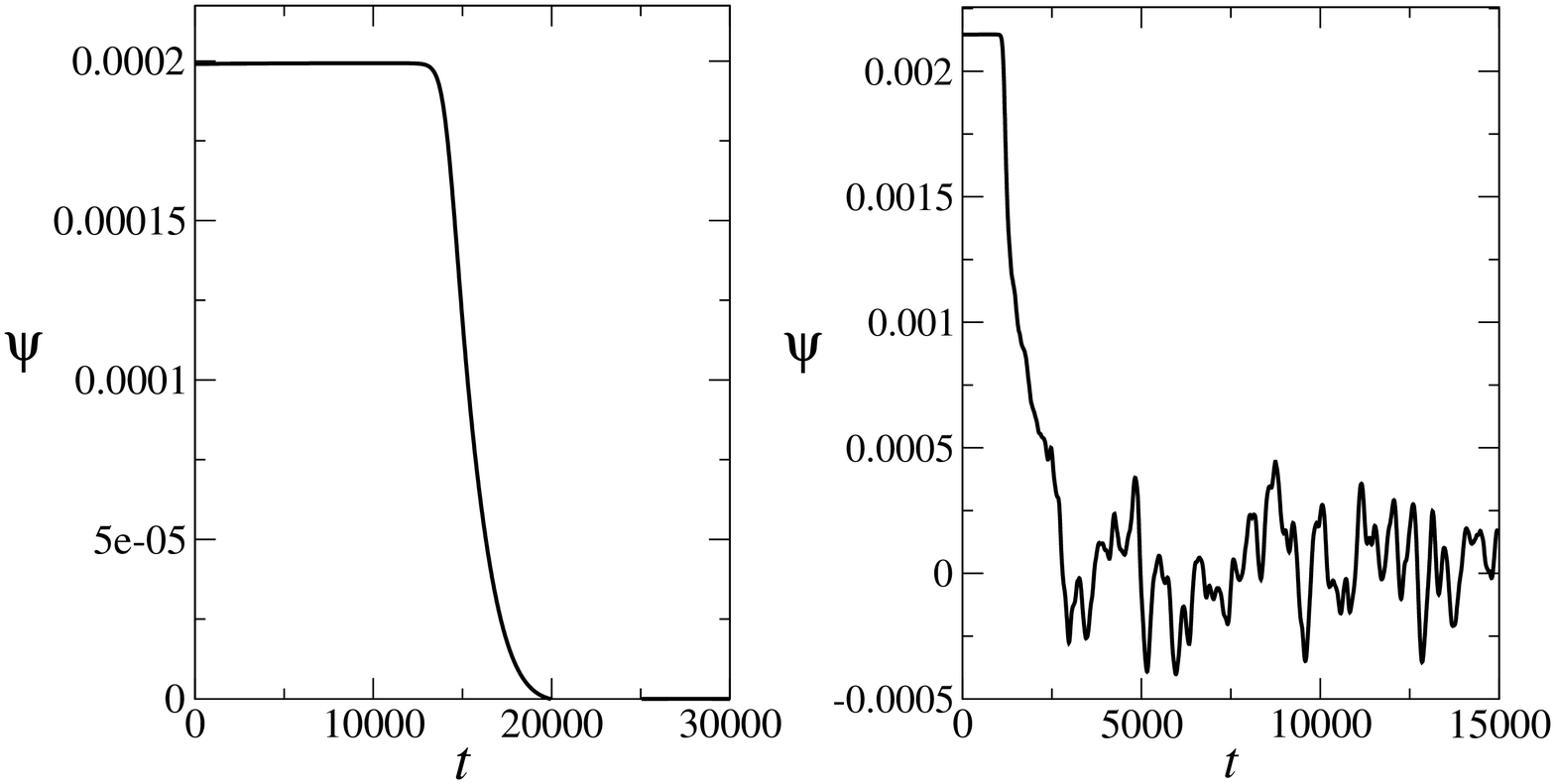}
  \caption{Throughput versus time for $l=0.002$ and $\zeta=0.004$
    (left), $\zeta=0.04$ (right).}
\label{fig:throughput}
\end{figure}

\begin{figure*}[tb]
  \includegraphics[width=8.5cm]{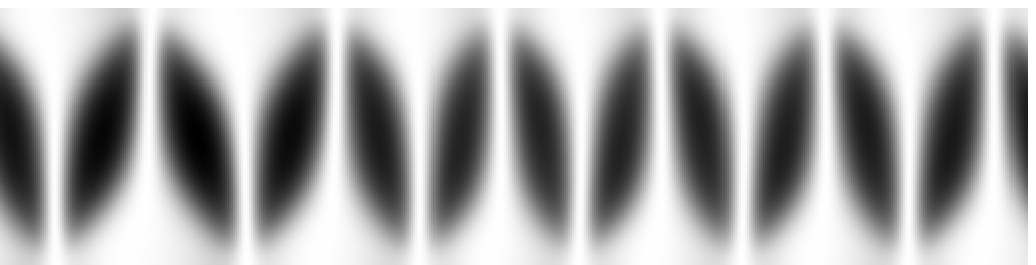}  
\includegraphics[scale=0.38,angle=90]{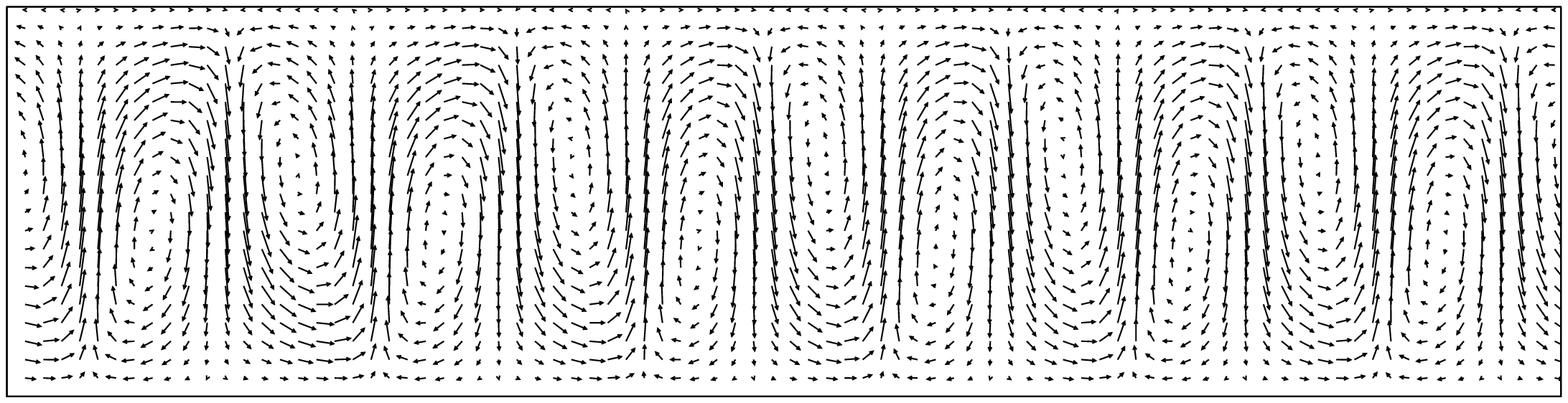}
  \includegraphics[width=8.5cm]{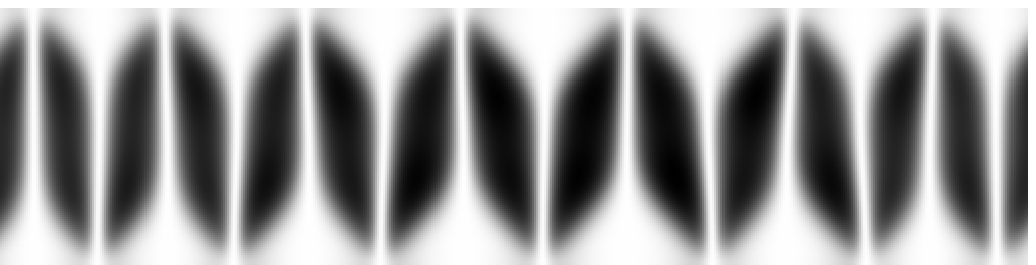}
  \includegraphics[scale=0.38,angle=90]{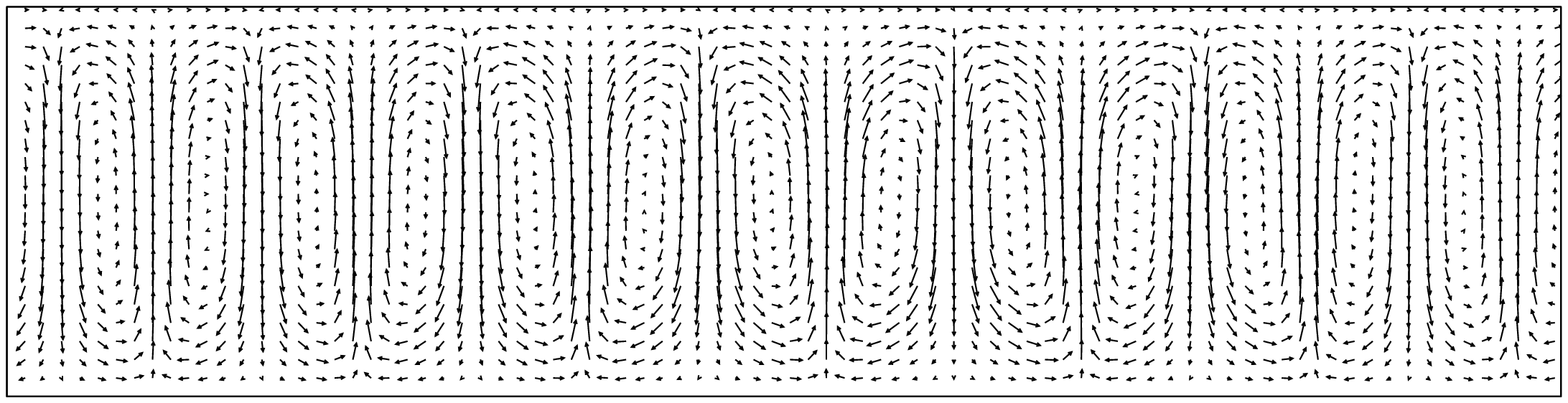}
  \includegraphics[width=8.5cm]{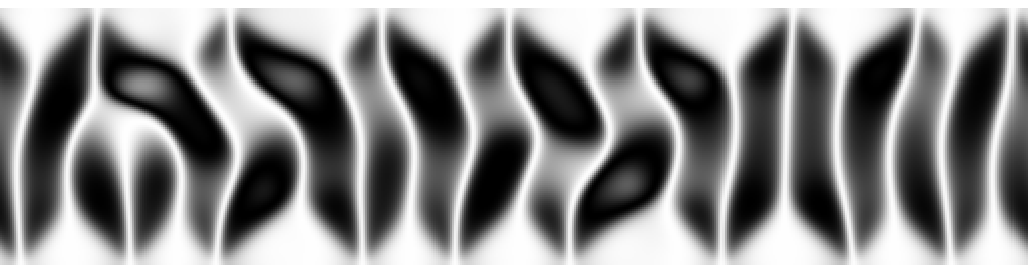}
  \includegraphics[scale=0.38,angle=90]{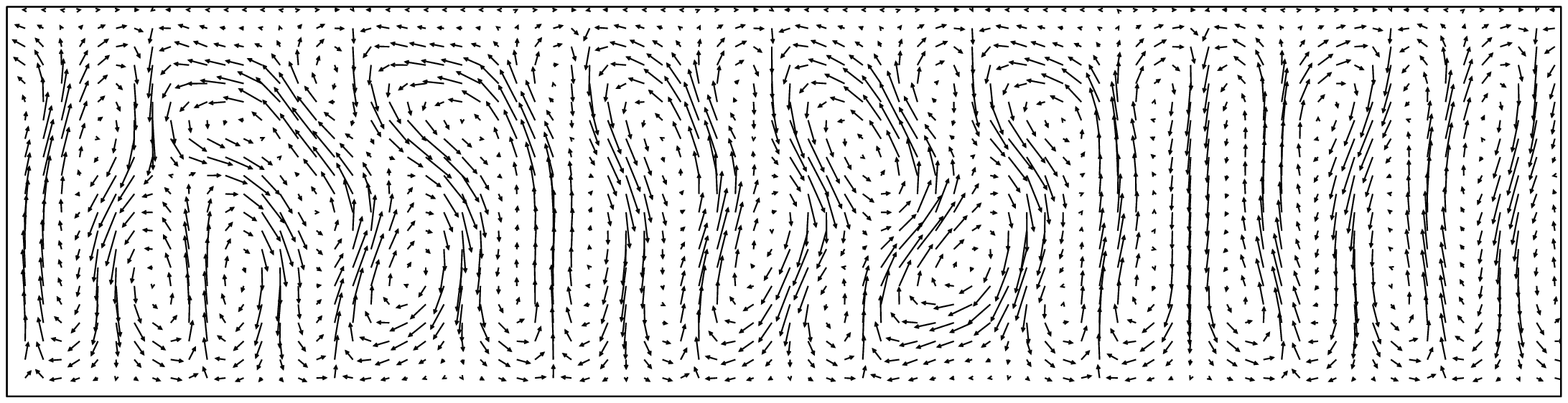}
  \includegraphics[width=8.5cm]{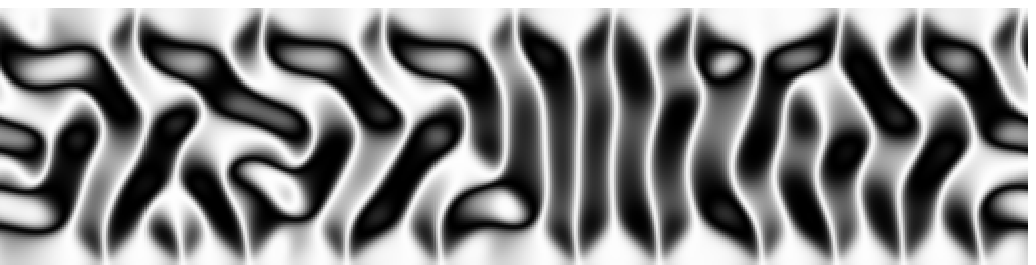}
  \includegraphics[scale=0.38, angle=90]{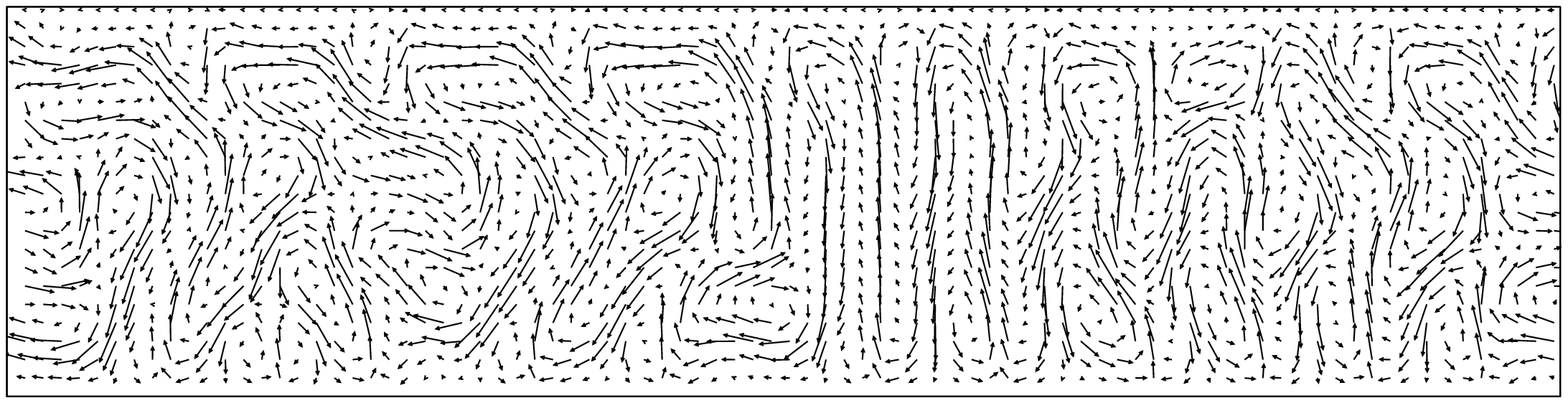}
  \includegraphics[width=8.5cm]{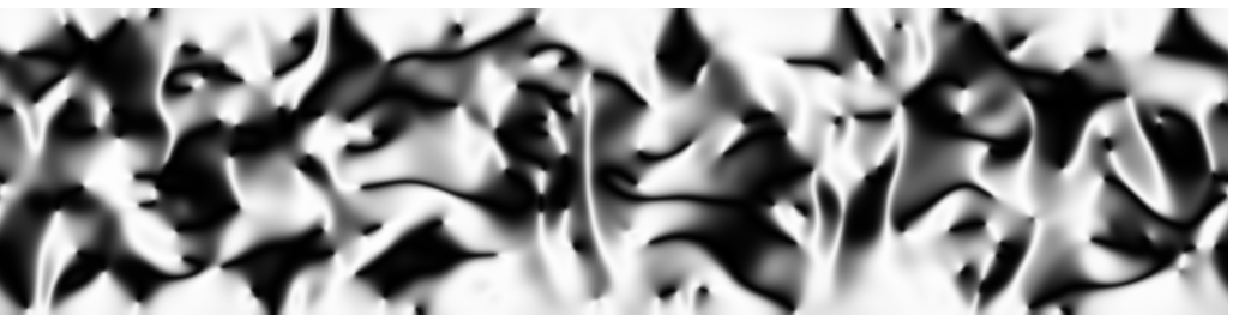}
  \includegraphics[scale=0.38,angle=90]{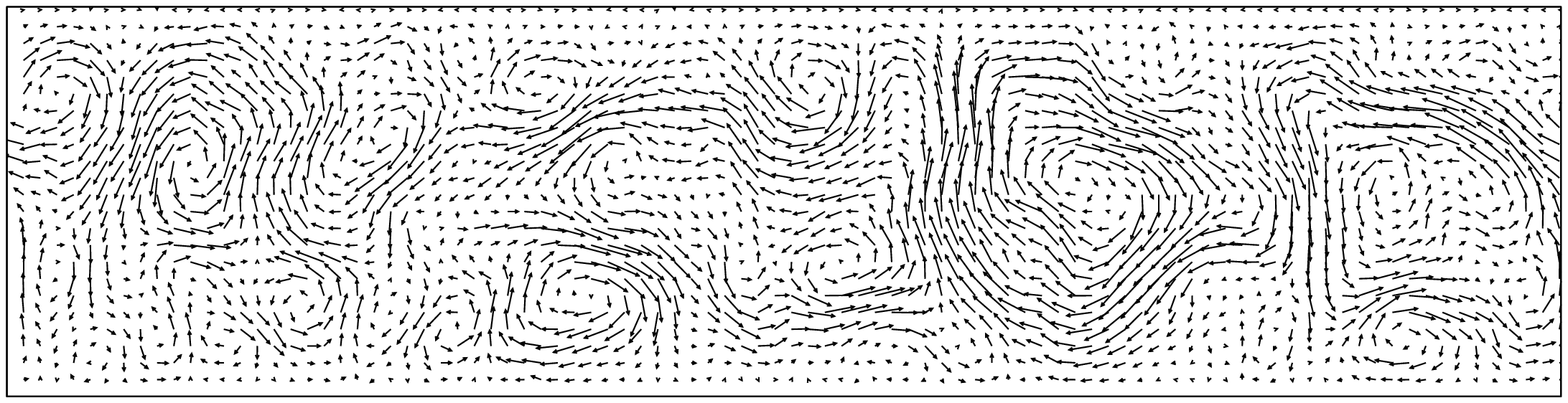}
  \includegraphics[width=8.5cm]{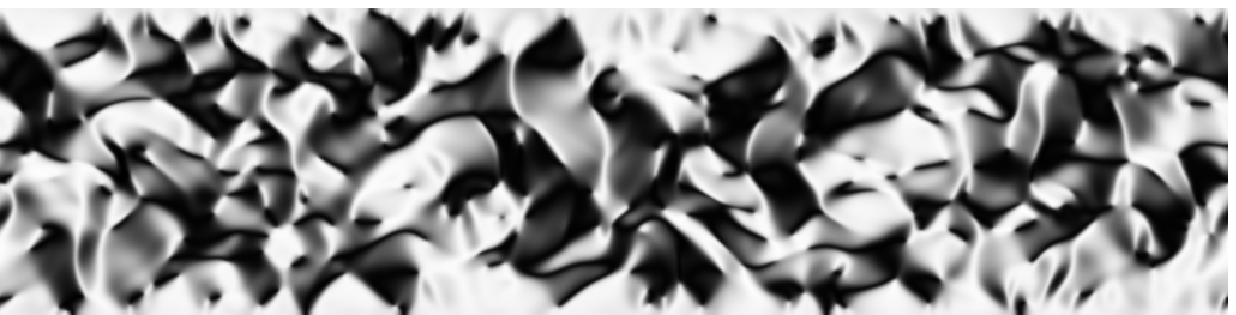}
  \includegraphics[scale=0.38,angle=90]{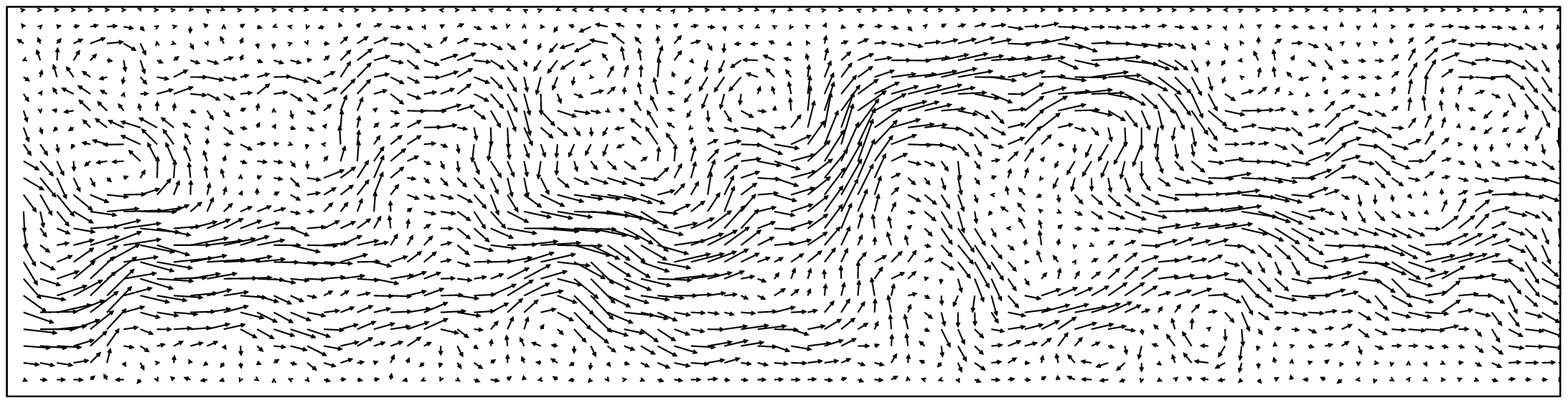}
  \caption{Snapshot at long time of 2D runs for
    $\zeta=0.002,0.004,0.01,0.02,0.04,0.1$ and
    $l=0.002$
(free boundary conditions, no external shear throughout). Greyscale on left shows $Q_{xx}$; while arrows on right show the fluid velocity. {The $x$ direction is horizontal, $y$ vertical.}}
\label{fig:states1}
\end{figure*}

\begin{figure*}[tb]
  \includegraphics[width=8.5cm]{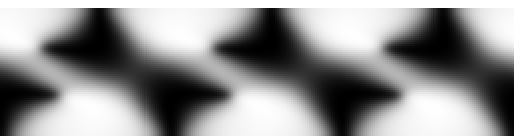}
  \includegraphics[scale=0.38,angle=90]{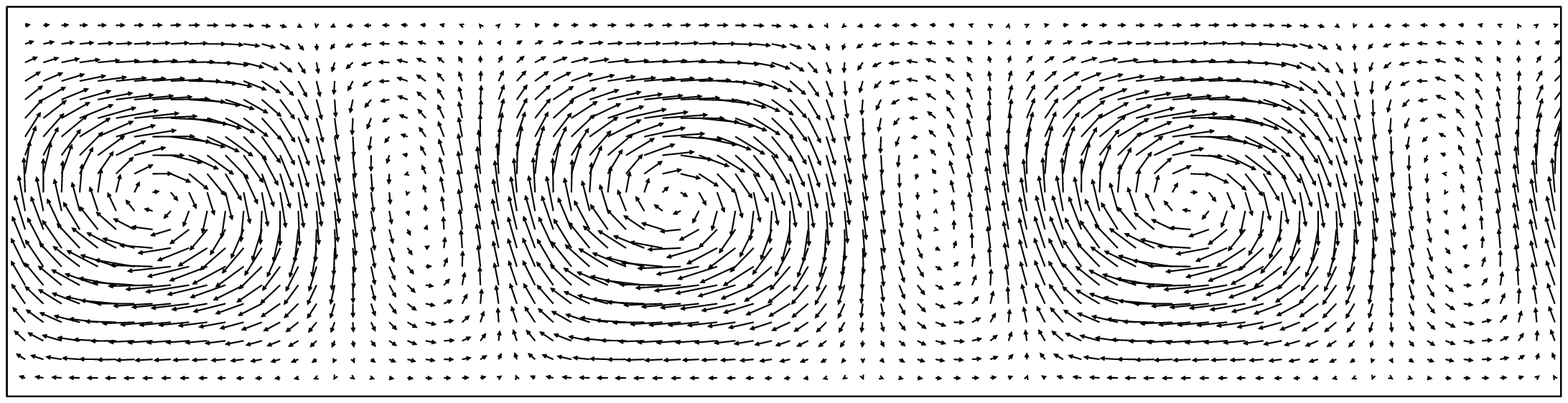}
  \includegraphics[width=8.5cm]{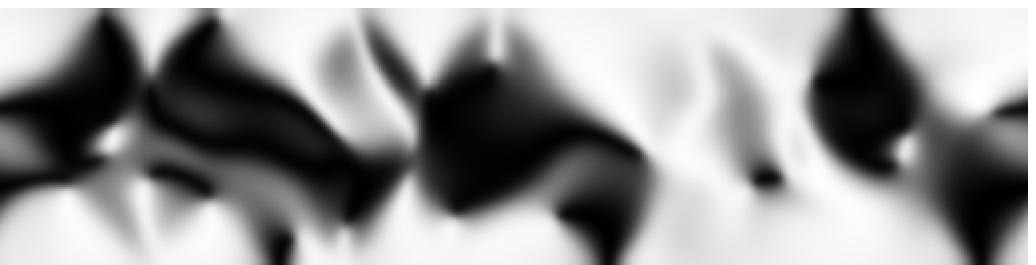}
  \includegraphics[scale=0.38,angle=90]{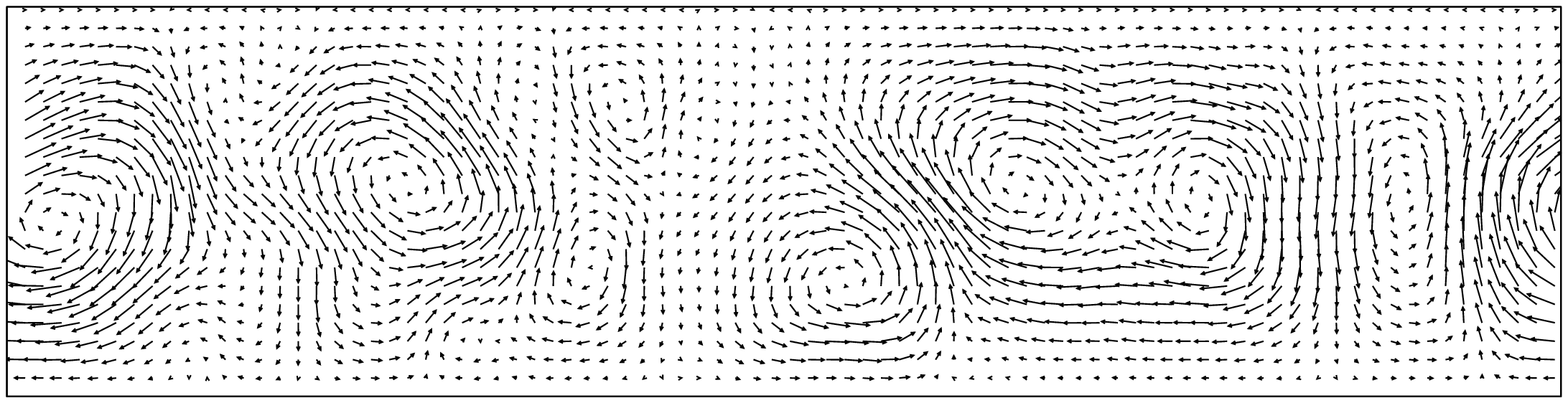}
  \includegraphics[width=8.5cm]{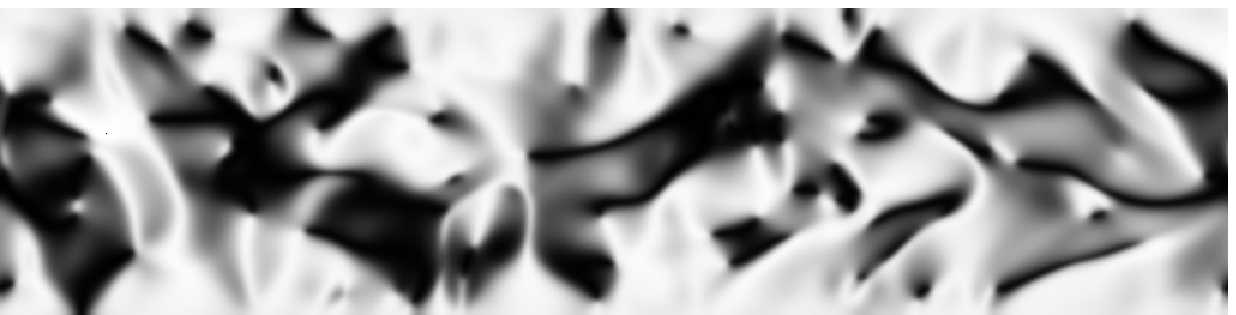}
  \includegraphics[scale=0.38,angle=90]{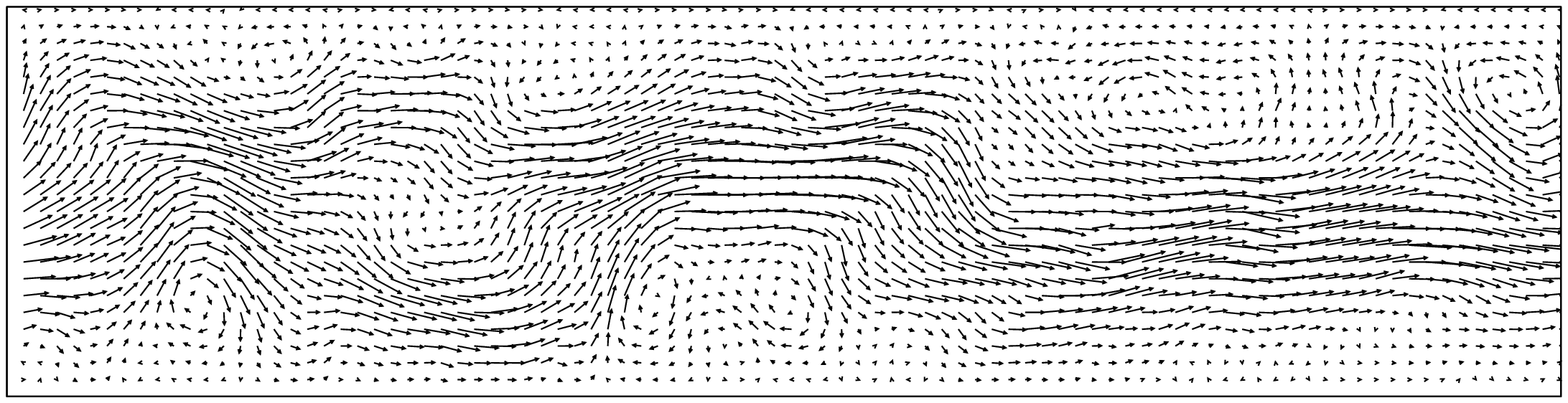}
  \includegraphics[width=8.5cm]{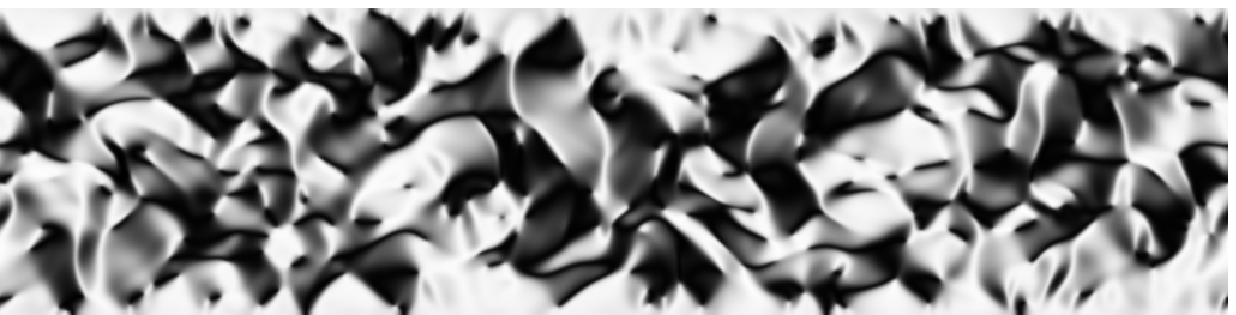}
  \includegraphics[scale=0.38,angle=90]{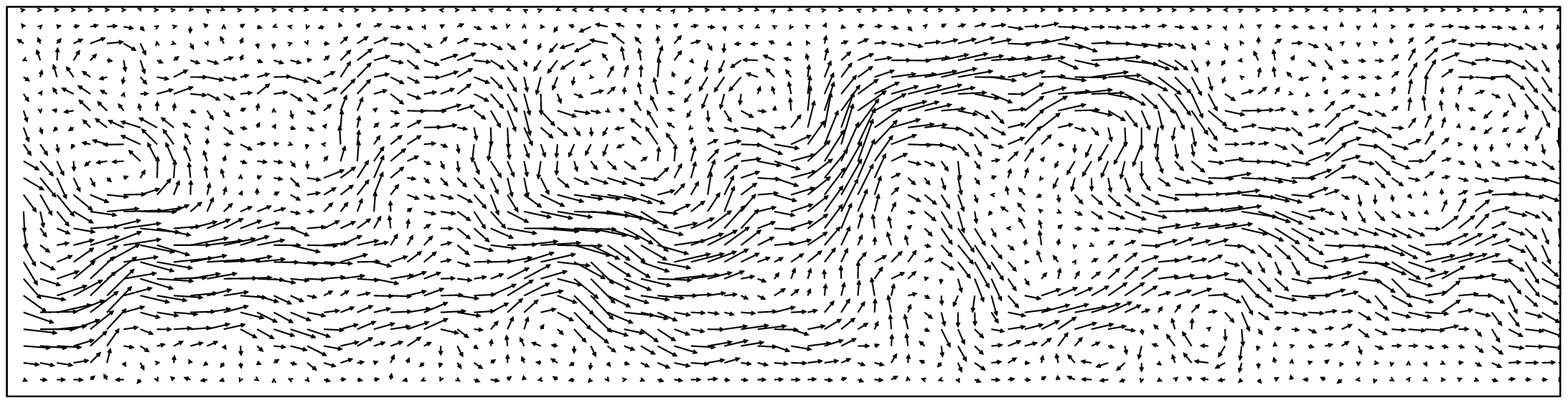}
  \caption{Snapshot at long time of 2D runs for $\zeta=0.1$ and,
from top to bottom, $l=0.016,0.008,0.004,0.002$ (free boundary conditions, no external shear throughout). Greyscale on left shows $Q_{xx}$; while arrows on right show the fluid velocity. {The $x$ direction is horizontal, $y$ vertical.}}
\label{fig:states2}
\end{figure*}

Using the free boundary condition, Eqn.~\ref{eqn:free}
above, we performed a series of 2D runs at the values of
$\zeta,l$ shown by squares in Fig.~\ref{fig:phaseDiagram2D}. In each
run we input as an initial condition the final state of the
corresponding 1D run of Sec.~\ref{sec:1D} above. (All these 1D runs
had themselves used free boundary conditions.) The dashed line in
Fig.~\ref{fig:phaseDiagram2D} is copied directly from
Fig.~\ref{fig:profiles1D}, bottom. Accordingly, all runs above
this line had a quiescent initial condition. All those below it
started with coexisting shear bands of equal and opposite shear rate.
To this 1D intial condition, we also added a 2D random component 
(the noise distribution was uniform, and could be either positive
or negative for each component of the order parameter tensor) of tiny
amplitude. This initialisation procedure allowed a study of the linear
regime kinetics for the initial growth (or decay) of 2D perturbations
at early times, as well as the ultimate attractor that is attained at
long times. 

For all runs located above the dashed line in
Fig.~\ref{fig:phaseDiagram2D}, the initial 2D perturbation decayed to
zero as a function of time, showing the homogeneous quiescent state to
be linearly stable.  In all runs below the line, for which the 1D base
state was shear banded, the initial 2D perturbation grew in time.  \if{The
dispersion relation for the initial growth away from the 1D base state
is shown in Fig.~\ref{fig:dispersion}. This was measured by tracking
the amplitude of the first few (lowest $q_x$) modes as a function of
time on a log-linear plot, of which the slope gives the growth rate
$\omega(q_x)$ for any given mode.}\fi

In each case, this destabilisation of the 1D initial state resulted at
long times in a more complicated state with 2D structure. A
macroscopic signature of this evolution is the decay of the
gap-averaged throughput, which was non-zero in the initial V-shaped 1D
velocity profile, to zero (on average) in the final 2D state; see
Fig.~\ref{fig:throughput}. (Note that by convention we choose the V-shaped velocity profile in the initial state 
to correspond to a positive throughput.)

At long times, the signal of $y-$integrated amplitude versus time (for
all $q_x$ modes in any given run) settled either to steady state, or
to an oscillatory attractor, or to an aperiodic, apparently chaotic, attractor. The word ``apparently'' is used here because we do not measure Lyaponov exponents and therefore do not distinguish true chaos from quasiperiodic motion. (From now on, though, we neglect this distinction and use the terms `chaotic' and `aperiodic' interchangeably.) We distinguish these three
different dynamical regimes by the type of filling of the squares in
Fig.~\ref{fig:phaseDiagram2D} (crosses=steady; shaded=oscillatory;
solid-filling=chaotic.)  Representative snapshots of the full 2D state
at a long time after the start of each run, once the system has
attained this ultimate attractor, are shown in Figs.~\ref{fig:states1}
and~\ref{fig:states2} respectively for runs performed along the
horizontal and vertical thin dotted lines in
Fig.~\ref{fig:phaseDiagram2D}. As can be seen, steady roll-like states
tend to dominate towards the upper left of the unstable regime in the
$\zeta,l$ plane, giving way to chaotic, turbulent-like states at the
lower right.

It is remarkable that the 1D shear bands are immediately unstable towards a 
static ``roll-like'' flow pattern with variation in both $x$ and $y$. 
Given this, 
however, the general progression from steady rolls via oscillations to chaotic 
flow on increasing activity at fixed $l$ (that is, fixed ratio of sample 
dimension to microscopic length, $L/l$) is fairly natural -- similar behavior 
was reported anecdotally for one parameter set in \cite{active_pre}. Notably, 
the same progression is seen at fixed $\zeta$ by instead decreasing $l$ -- 
equivalent to increasing $L/l$, the ratio of the sample size to the molecular 
length scale. Thus not only is the instability of the quiescent state delayed 
in small systems (as it is in 1D, \cite{Cates08}), but also the subsequent 
transitions to oscillation and chaos are likewise delayed. This reflects the 
high energy cost of creating inhomogeneous director fields in systems that are 
not many times larger than the elasticity length $l$; this cost stabilizes the 
quiescent state but can be overcome, for all the $l$ values studied here, by 
increasing the activity sufficiently.

\subsubsection{Fixed boundary conditions}

\begin{figure}[tb]
  \includegraphics*[width=7.5cm]{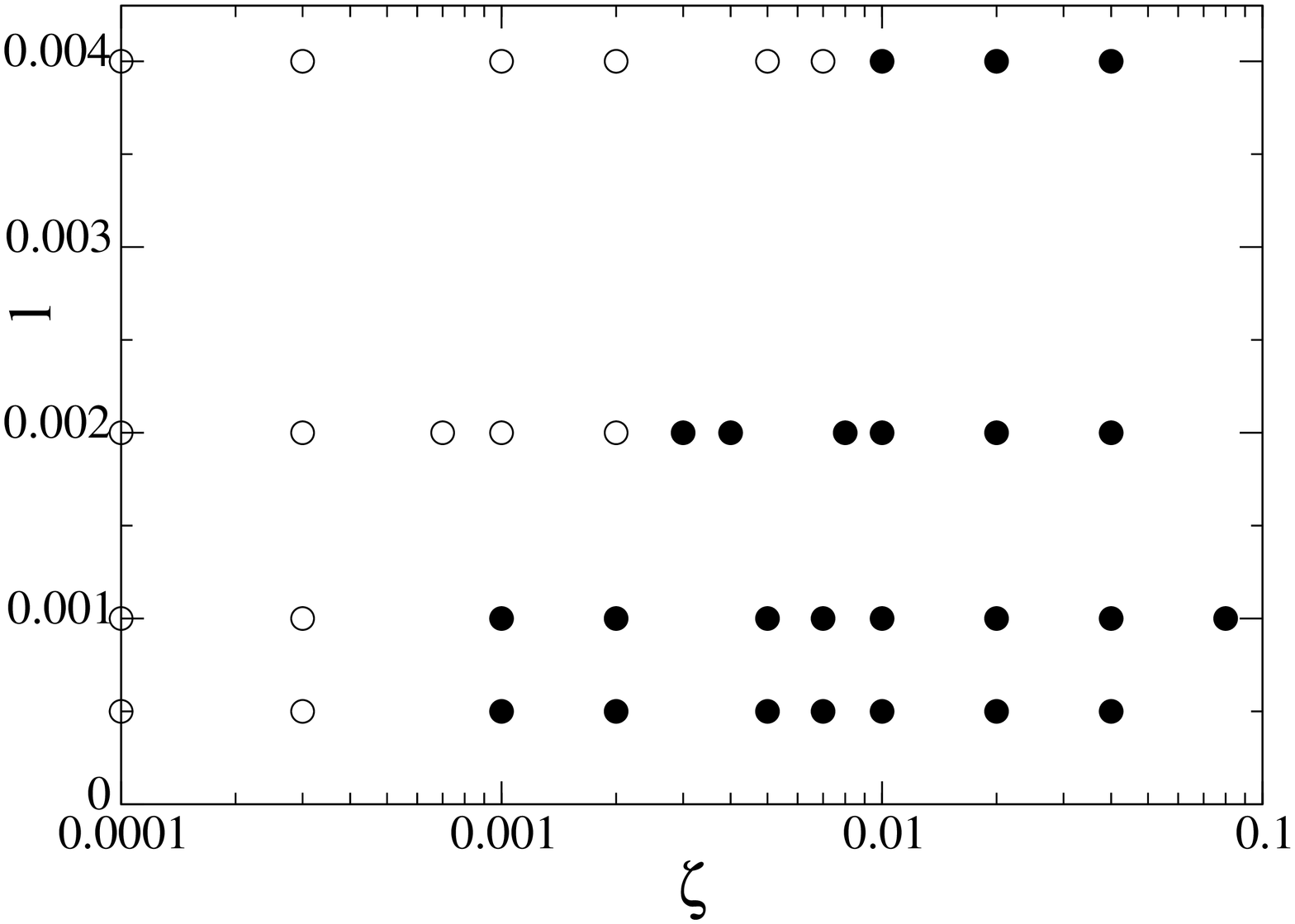}
  \includegraphics*[width=7.5cm]{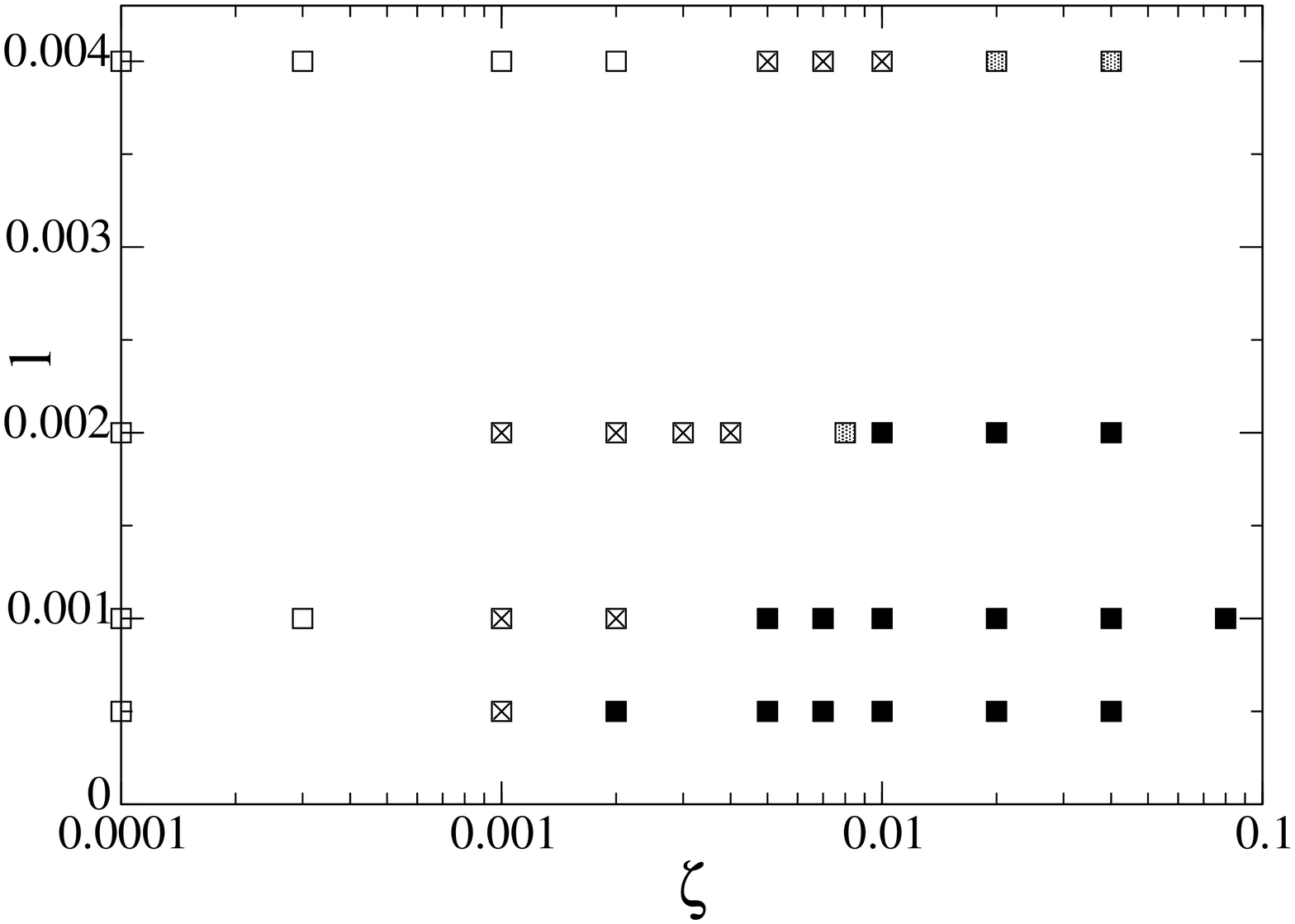}
  \caption{Top: Phase diagram for 1D runs with fixed 
boundary conditions, each denoted by a circle. Empty circles denote
quiescence, filled ones spontaneously flowing states.
Bottom: Phase diagram for 2D runs with fixed boundary conditions,
each denoted by a square. Empty
    squares: quiescence. Elsewhere the stable 1D profile with a small 
inhomogeneity gives
    way to a state of 2D domains that is: steady (crosses);
    oscillatory (shaded square); chaotic (filled squares).}
\label{fig:phaseDiagram2Dfixed}
\end{figure}

We now turn to the case of fixed boundary conditions. 
Fig.\ref{fig:phaseDiagram2Dfixed} shows the phase diagram in this case, 
plotted in the same way as Fig.\ref{fig:phaseDiagram2D} for free BCs. Just as 
we found there, the simple 1D banded state (creating a V-shaped velocity 
profile) is never seen.
As expected, in the regime of small $l$ (large $L/l$) the BCs become relatively
unimportant and the behavior as a function of $\zeta$ very similar to that 
with free BCs. However, at larger $l$ (i.e., smaller system sizes in units of 
the microscopic length), the choice of BCs plays a larger role.

In 1D the fixed BCs, by constraining the director to lie in the flow plane, 
somewhat suppress the instability towards the banded state relative to free 
BCs \cite{Cates08}. However, the 1D banded state is again unstable; moreover, 
the presence of additional instability modes pushes the boundary of the 
passive phase back towards lower activity.

\begin{figure*}
\begin{center}
\centerline{\includegraphics[scale=0.5]{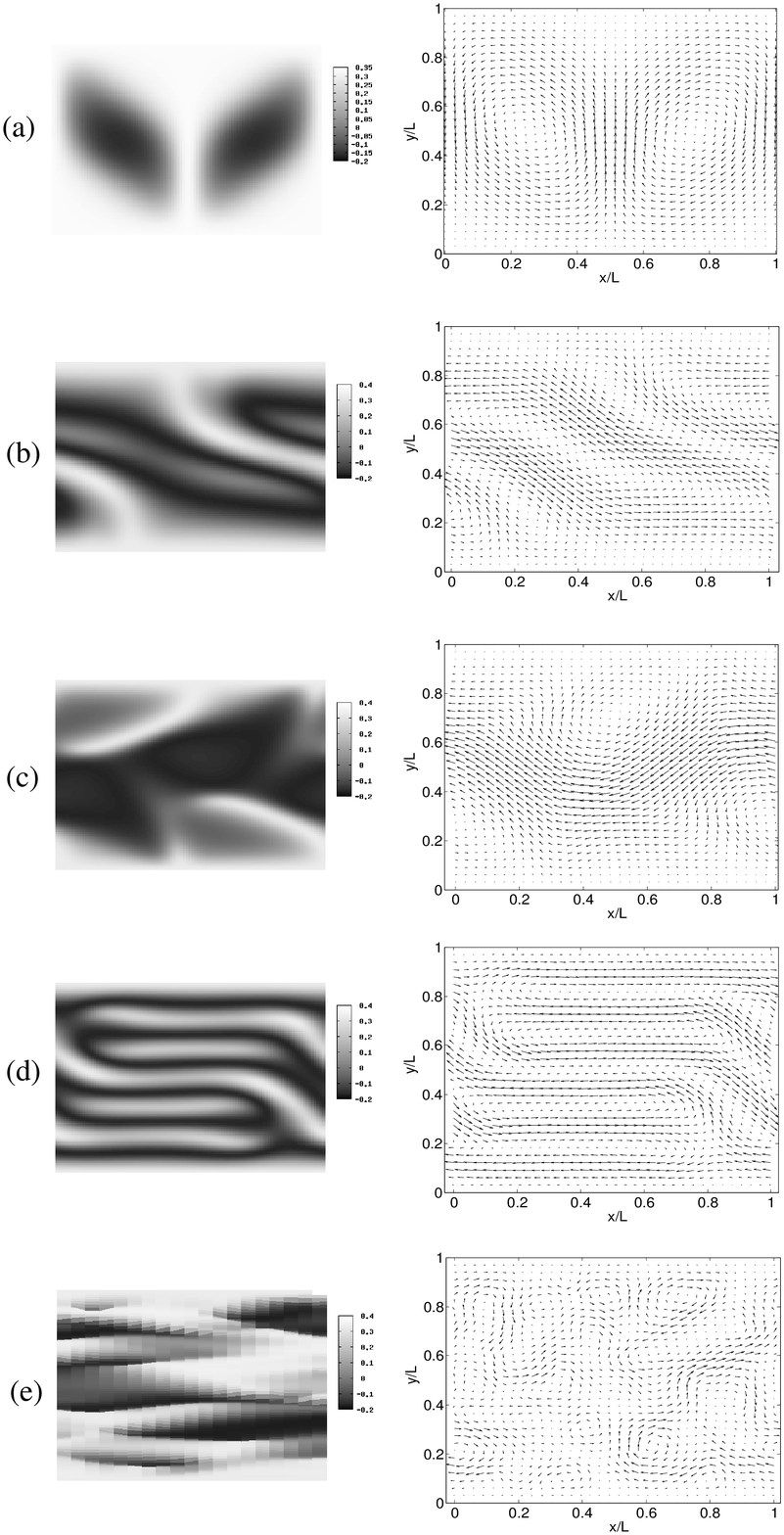}}
\end{center}
\caption{Snapshots corresponding to 
simulations with extensile fluids and fixed boundary conditions,
for a system with $L_x=L_y=1$ and periodic boundary conditions along the
flow direction. 
The snapshots show grayscale plots of $Q_{xx}$ (left column), and
the velocity profile (right column).
The value of the activity is (from top to bottom): 0.002 (a), 0.004 (b), 0.008 (c), 0.01 (d), 0.08 (e); throughout, $l=0.002$.}
\label{unsheared_fixedBC}
\end{figure*}

Specifically, to numerical accuracy we discern the following progression at 
fixed $l=0.002$. If the activity is smaller than, but close to, the critical threshold in 1D,
the system forms steady convection rolls (seen previously in \cite{active_pre}). Fig. \ref{unsheared_fixedBC}a gives a steady-state snapshot of $Q_{xx}$ and of
the fluid flow in this regime. This steady state is achieved in about 200,000 LB timesteps which corresponds to time $6.755 \times 10^4$  in our units 
(i.e., $t=6.755 \times 10^{4} \tau$).
On the other hand, if the activity is raised to just beyond the level needed to create spontaneous flow
in 1D, then the rolls break and form what looks like one or more tilted
bands (Fig. \ref{unsheared_fixedBC}b). These tilted bands are again stationary.
Additional simulations in a box with a 4:1 aspect ratio (not shown) suggest that these
may be stabilised by the reduced geometry and box size used here -- 
consistently with the fact that we do not
observe these states with free boundary conditions in a 4:1 box (Section VA1). 

If the activity increases further (Fig. \ref{unsheared_fixedBC}c, $\zeta = 0.008$) a roll-like state that forms initially breaks into an apparently undulating band which then oscillates.
This looks quite similar to Fig. \ref{fig:states2} (second panel from top) which is however already chaotic (perhaps because of the larger aspect ratio).
A further increase to $\zeta = 0.01$ appears to stabilise multiple bands
with small vortices in between (Fig. \ref{unsheared_fixedBC}d). These states are quasi-1D in
nature, in the sense that the main variation occurs along the velocity gradient
direction, yet were not seen in our strictly 1D simulations.
For yet larger activity, we have an apparently chaotic flowing state, similar to that reported above for 
free BCs (Fig.\ref{unsheared_fixedBC}e), also seen previously in Ref. \cite{active_pre}. 

It should be noted that the threshold for spontaneous flow in 1D and 2D are different with fixed boundary conditions. This is a difference with respect to the free boundary condition results discussed before, which is due to boundary stabilisation of the passive phase. For relatively large values of $l/L$, the boundaries provide layers within which the ordering is fixed by the anchoring, and this provides the observed stabilisation. As expected, we find that as the value of
$l/L$ decreases, so does the difference between the thresholds in 1D and 2D (see Fig.~\ref{fig:phaseDiagram2Dfixed}).


\subsection{Sheared systems}
\label{sec:sheared}

Here we address systems that are subject to a shear flow applied
externally at the cell boundaries. As before we consider in turn the cases
of free and fixed boundary conditions.

\subsubsection{Free boundary conditions}

For very small $\gdotbar$, one expects the effect of flow to be perturbative 
on the underlying zero-shear states found above. In a strictly 1D flow showing 
shear bands, the effect is merely to shift the band interface, creating a 
nonzero mean shear rate without altering any structure within the bands. This 
degeneracy allows the system to accommodate a small macroscopic shear flow 
without developing a stress (called ``superfluidity'' in \cite{Cates08}). 
In 1D we also found that increasing the shear rate beyond the superfluidity 
window caused elimination of one of the two shear bands, thereby restoring a 
homogeneous shear flow.

In 2D, with a more complicated inhomogeneous flow in the absence of
bulk shear, it is far from clear whether a similar degeneracy ought to
exist. If it does not, one expects a finite shear rate to be
accompanied by a finite stress, so that the superfluid region is
replaced by one of finite (zero-shear) viscosity. Nonetheless one
might expect restoration of homogeneous flow at high enough
$\gdotbar$, either by a sharp transition or only gradually as
$\gdotbar\to\infty$.

To address the behavior in 2D we use for reference the unsheared phase
diagram in the $(\zeta,l)$ plane of Fig.~\ref{fig:phaseDiagram2D},
focusing on three locations in it: $l=0.002$ and
$\zeta=0.001,0.005,0.04$, for which the unsheared state comprised
simple rolls, wavy rolls and turbulence respectively.  (Recall
Fig.~\ref{fig:states1}.) At each of these locations we perform a
series of runs for increasing values of the applied shear rate
$\gdotbar$. At each shear rate separately we perform shear startup
from a state of no flow, with an initial condition in which the order
parameter tensor is taken to be everywhere uniaxial for simplicity,
$Q_{\alpha\beta}=\tfrac{3}{2} S_1 (n_\alpha n_\beta -
\delta_{\alpha\beta}/3)$, and with its principal axis assumed to
reside in the $xy$ plane. To avoid biasing the system into any
particular final state, at each individual grid point separately we
assign a random value for the order parameter $S_1$ (drawn from a flat
top-hat distribution); and for the angle of the principal axis (from a
flat distribution between $0$ and $2\pi$).

\begin{figure}[tbp]
\includegraphics[width=7.5cm]{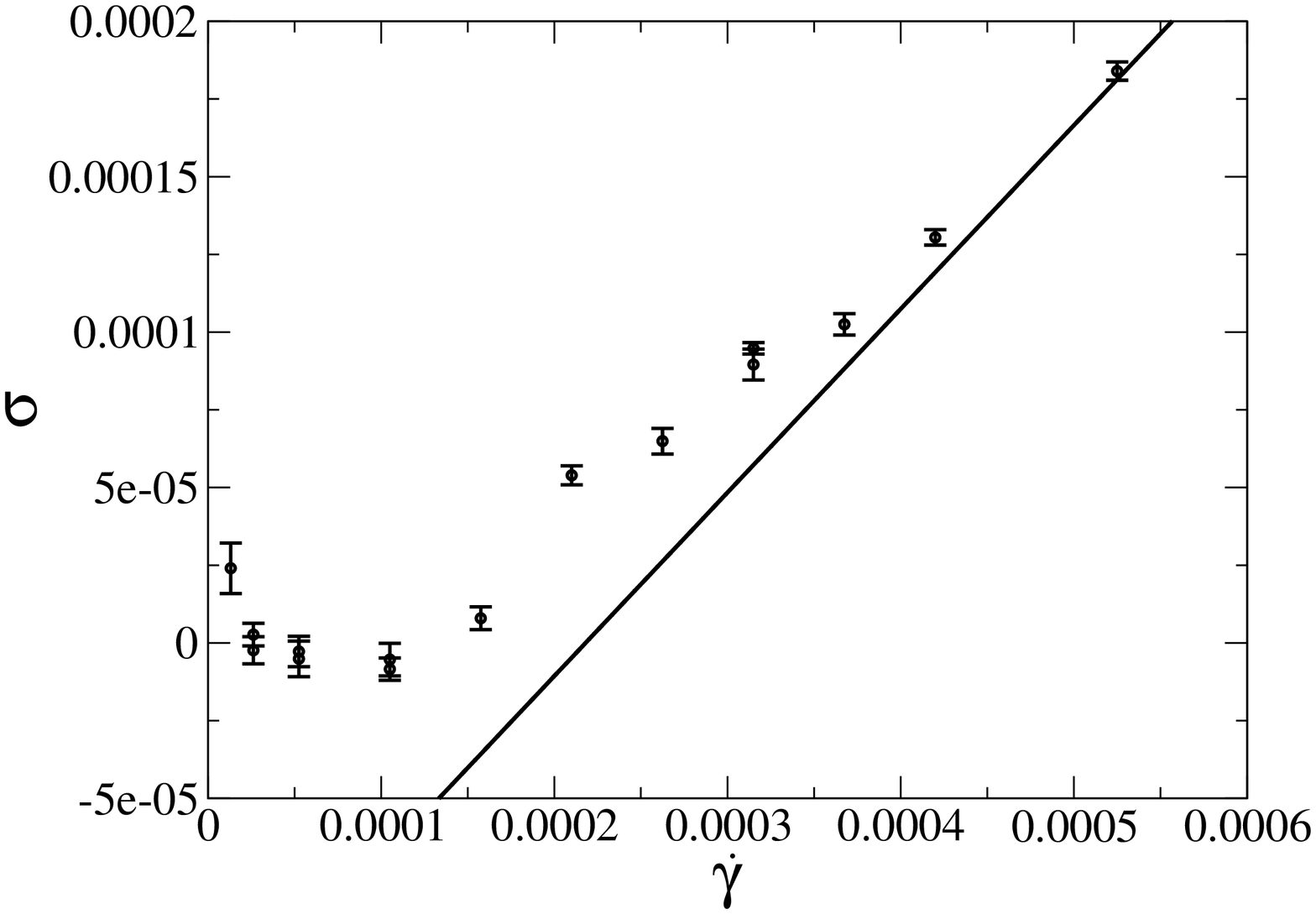}
\includegraphics*[width=6.5cm]{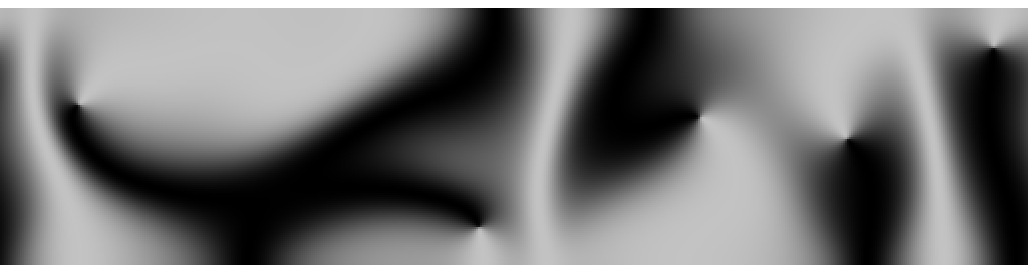}
\includegraphics*[width=6.5cm]{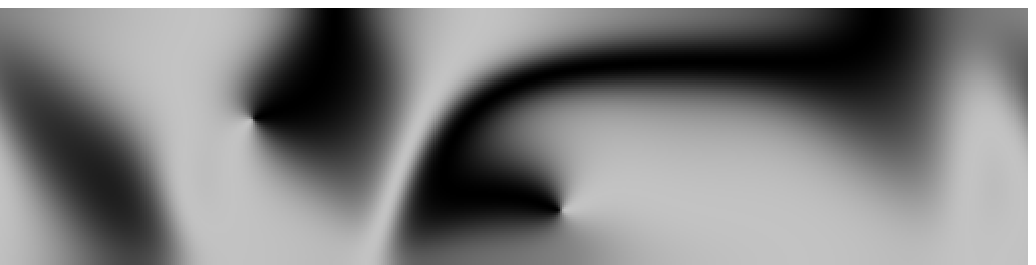}
\includegraphics*[width=6.5cm]{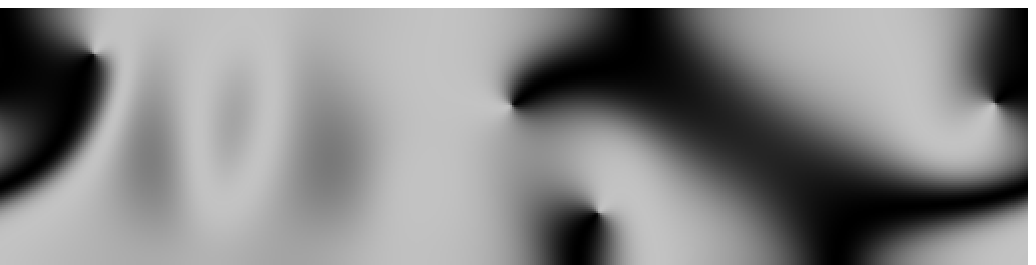}
\includegraphics*[width=6.5cm]{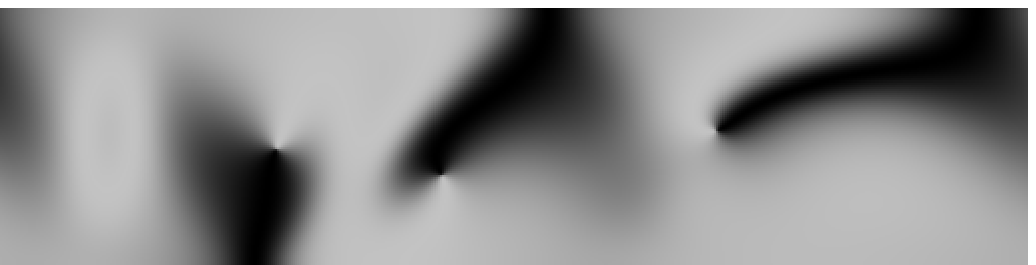}
\includegraphics*[width=6.5cm]{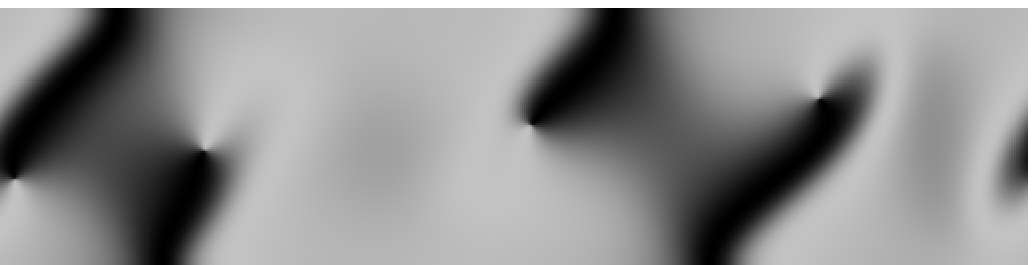}
\includegraphics*[width=6.5cm]{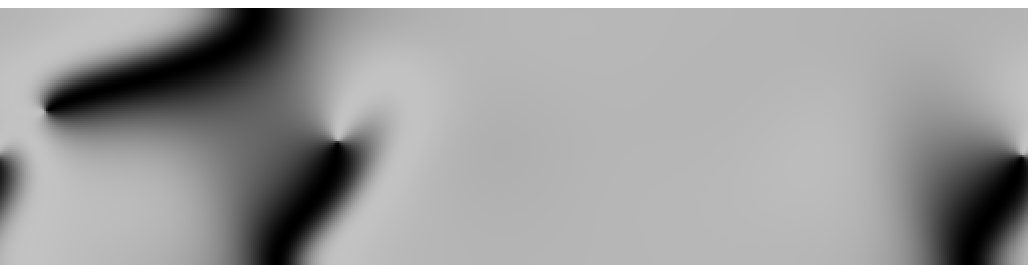}
\includegraphics*[width=6.5cm]{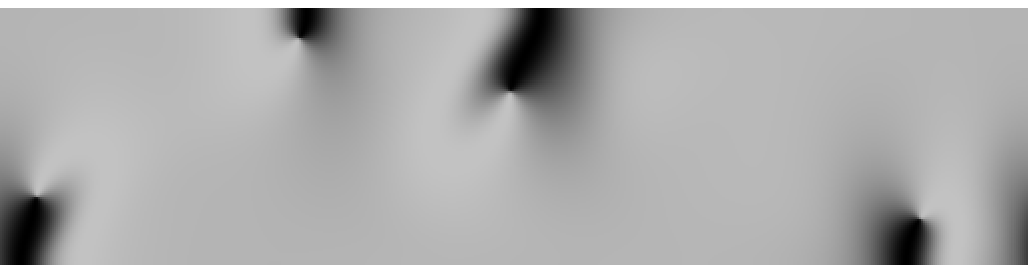}
\caption{Top: Flow curve for $\zeta=0.001, l=0.002$. {Lower:} representative state snapshot at a late time
  (on the final attractor) for each of the lowest seven shear
  rates. (Shear rate increasing in snapshots downwards. {The $x$ direction is horizontal, $y$ vertical.})
The solid line shows the 0D homogeneous constitutive curve.}
\label{fig:zeta0.001_l0.002}
\end{figure}

\begin{figure}[tbp]
\includegraphics*[width=7.5cm]{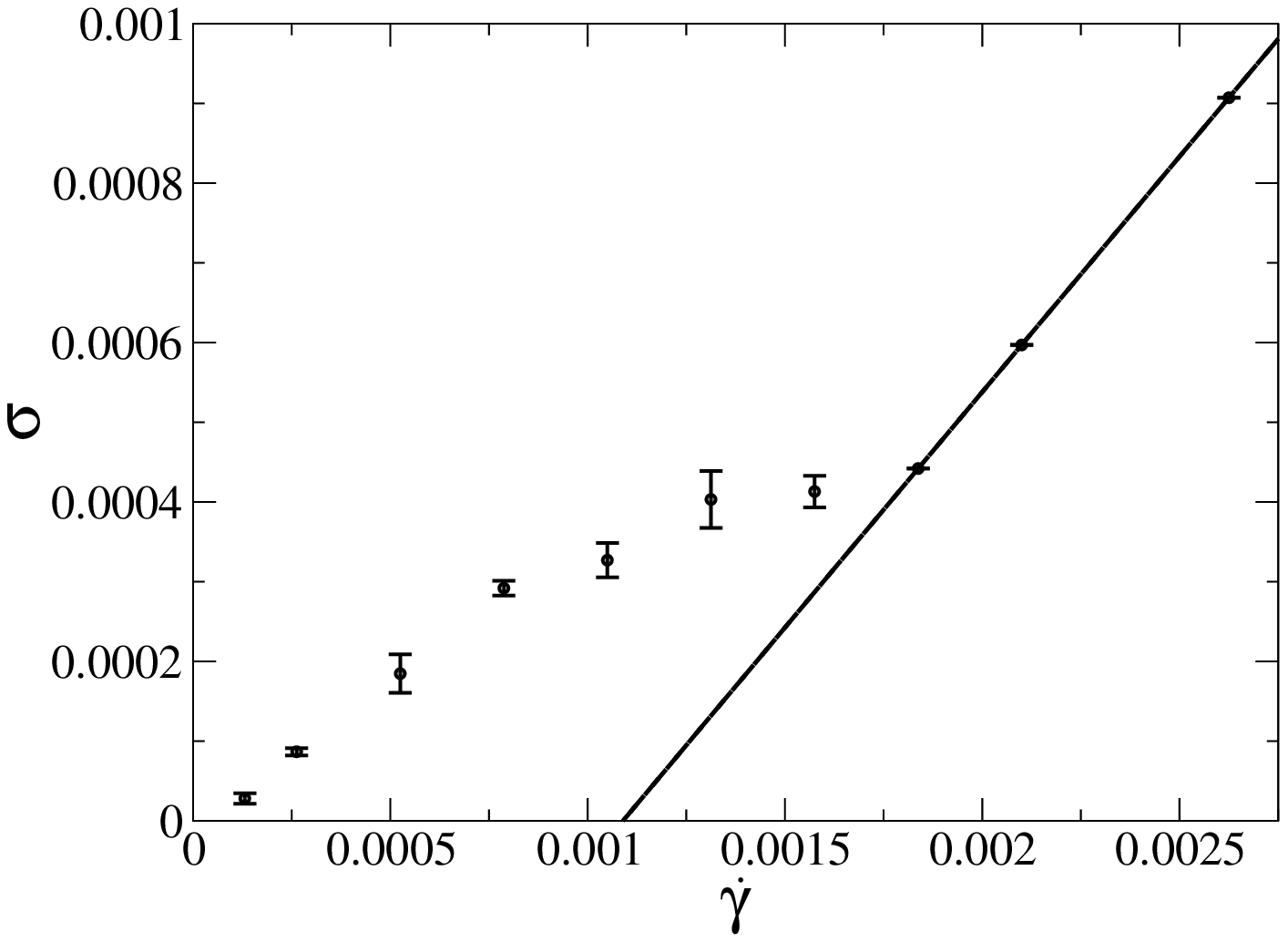}
\includegraphics*[width=6.5cm]{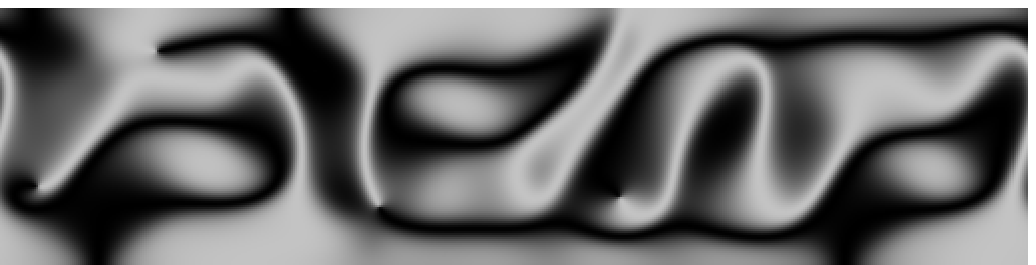}
\includegraphics*[width=6.5cm]{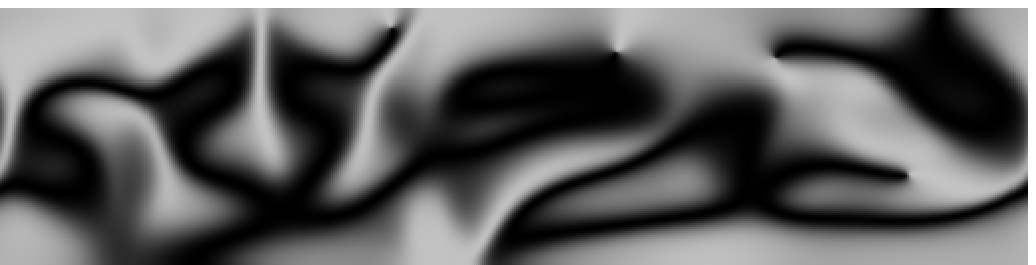}
\includegraphics*[width=6.5cm]{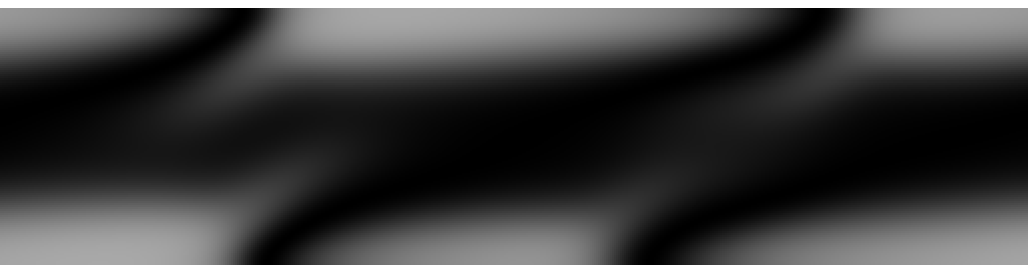}
\includegraphics*[width=6.5cm]{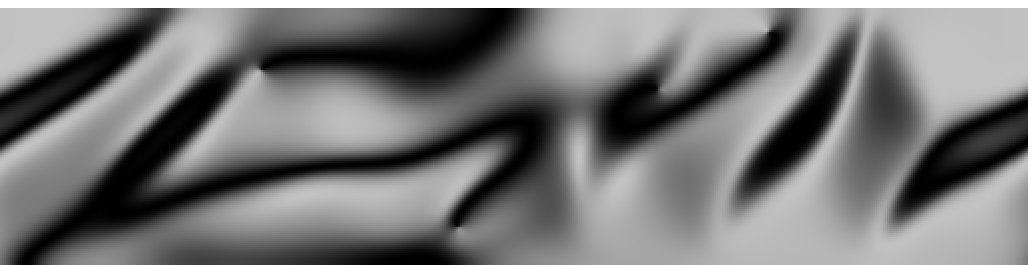}
\includegraphics*[width=6.5cm]{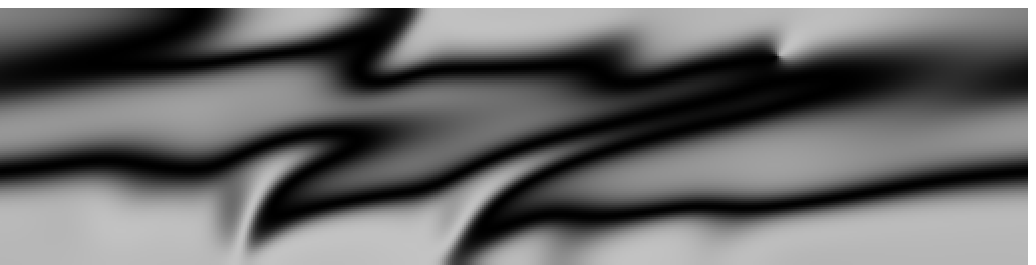}
\includegraphics*[width=6.5cm]{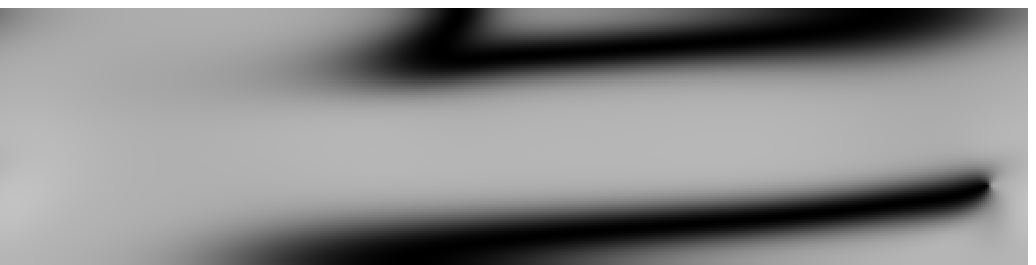}
\includegraphics*[width=6.5cm]{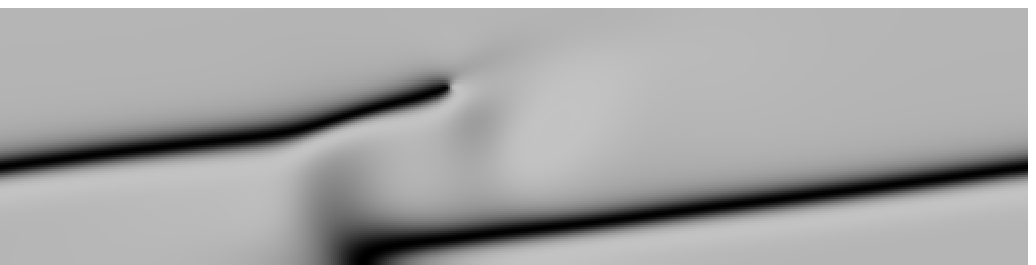}
\caption{Top: Flow curve for $\zeta=0.005, l=0.002$.   {Lower:}
  representative state snapshot at a late time (on the final
  attractor) for each of the lowest seven shear rates. (Shear rate
  increasing in snapshots downwards. {The $x$ direction is horizontal, $y$ vertical.})
The solid line shows the 0D homogeneous constitutive curve.}
\label{fig:zeta0.005_l0.002}
\end{figure}

\begin{figure}[tbp]
\includegraphics*[width=7.5cm]{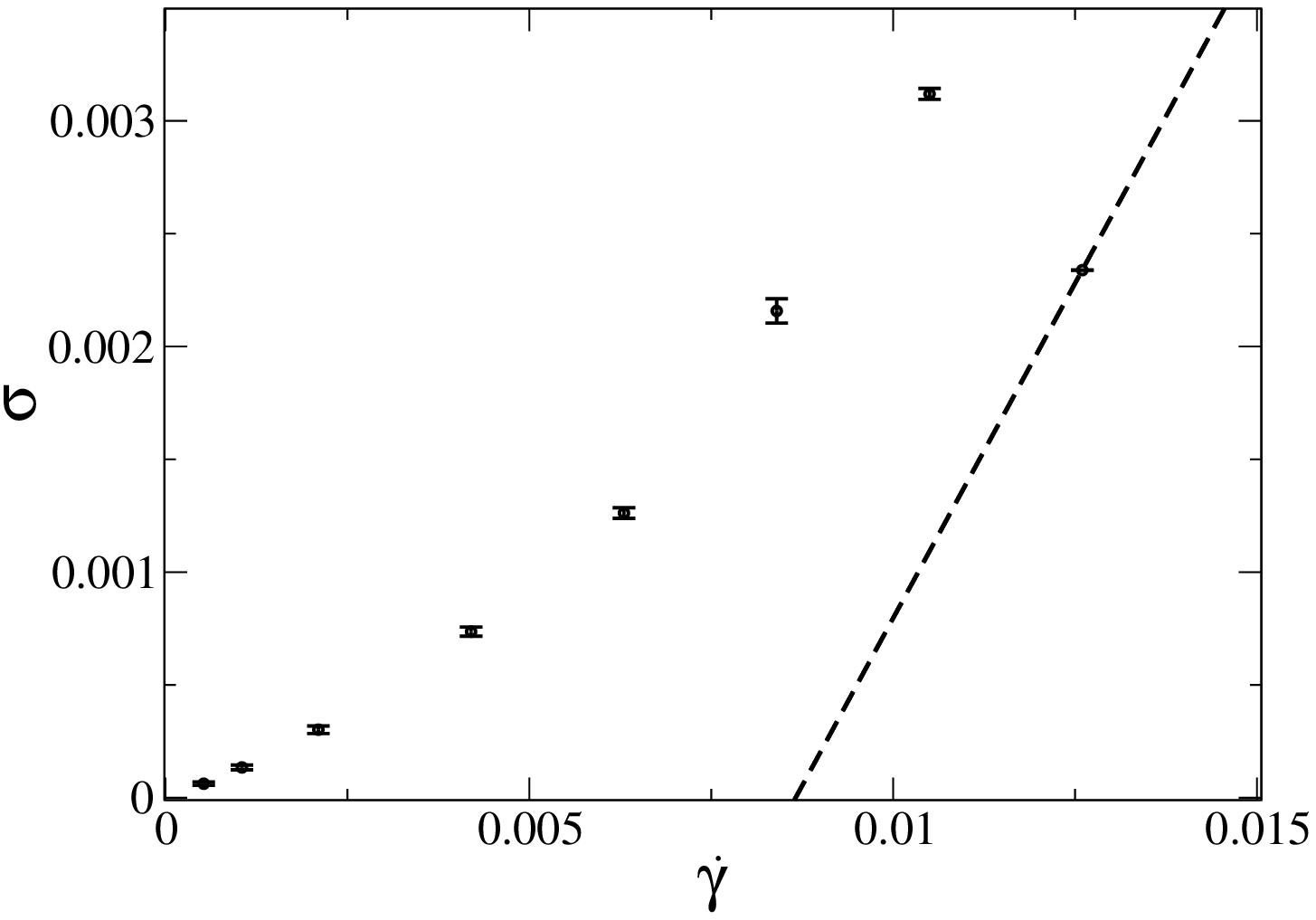}
\includegraphics*[width=6.5cm]{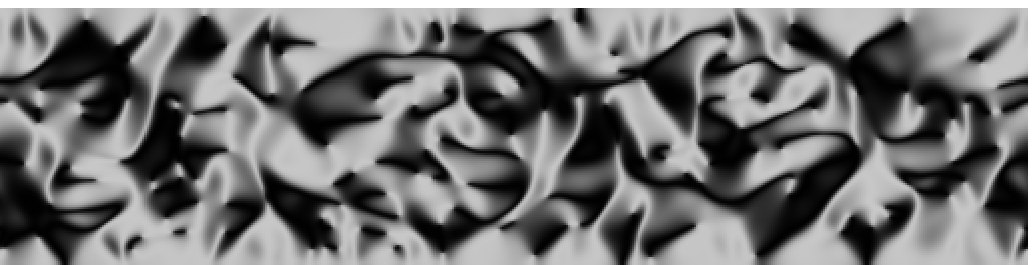}
\includegraphics*[width=6.5cm]{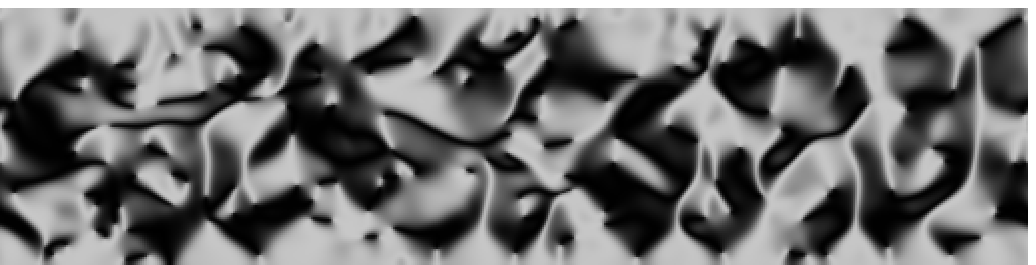}
\includegraphics*[width=6.5cm]{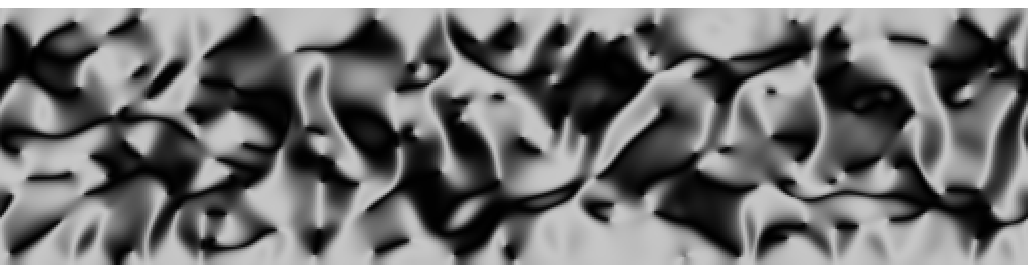}
\includegraphics*[width=6.5cm]{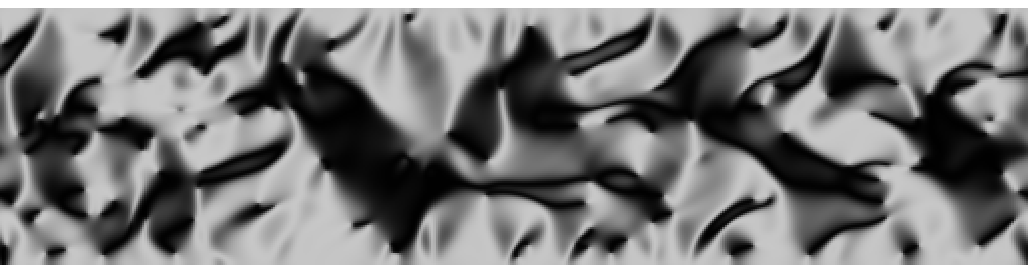}
\includegraphics*[width=6.5cm]{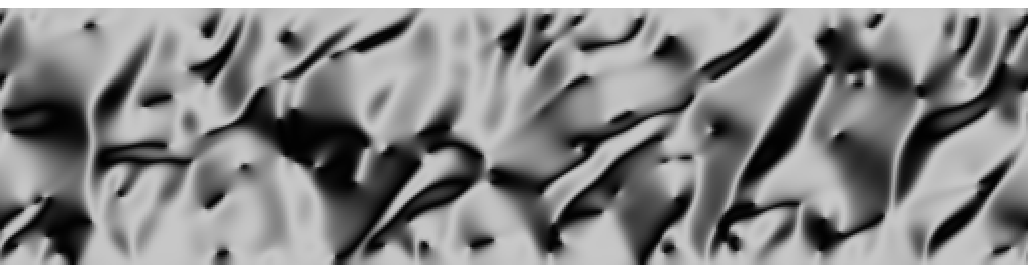}
\includegraphics*[width=6.5cm]{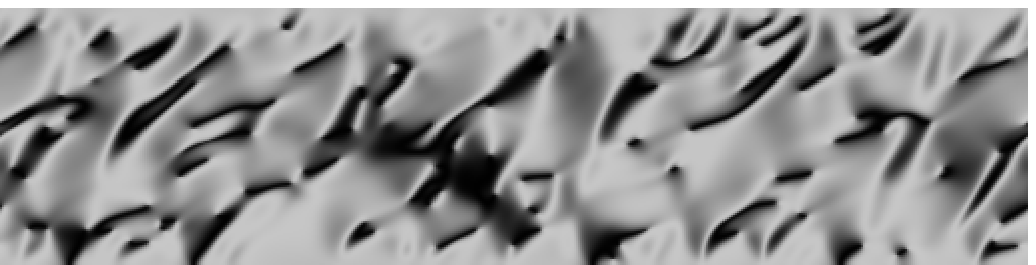}
\includegraphics*[width=6.5cm]{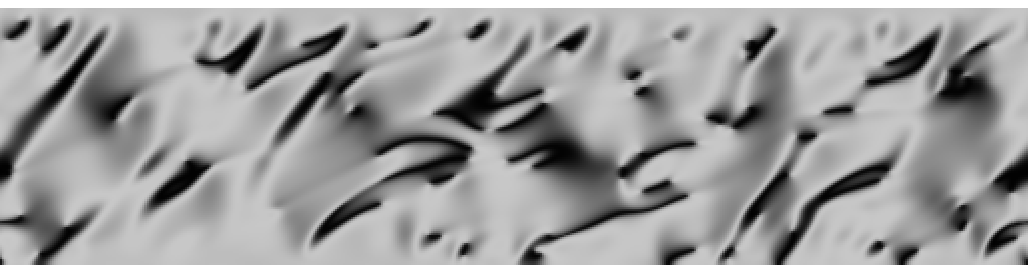}
\caption{Top: Flow curve for $\zeta=0.04, l=0.002$.   {Lower:}
  representative state snapshot at a late time (on the final
  attractor) for each of the lowest seven shear rates. (Shear rate
  increasing snapshots downwards. {The $x$ direction is horizontal, $y$ vertical.})
The solid line shows the 0D homogeneous constitutive curve.}
\label{fig:zeta0.04_l0.002}
\end{figure}

Our results are shown in Figs.~\ref{fig:zeta0.001_l0.002}
to~\ref{fig:zeta0.04_l0.002}.  For low activity $\zeta = 0.001$, the
static roll pattern observed without shear is destroyed even at the
smallest shear rates studied, whereas at high enough shear rates a
homogeneous flow is recovered and the shear stress reverts to the one
calculated in 1D. The flow pattern at very low shear rates is unsteady
and is spatially nonuniform with no evident periodicity in either
space or time. The shear stress takes both positive and negative
values. The unsteadiness makes it very difficult to determine a
reliable value for the average shear stress at small $\gdotbar$ (error
bars in Fig.~\ref{fig:zeta0.001_l0.002} are standard deviations of the mean
from the time series). 

At $\zeta = 0.005$, the `wavy roll' state observed in the case of zero
bulk shear (closely resembling that of Fig.~\ref{fig:states1}) is again
broken up at the smallest shear rates studied. This is not surprising
since this state has some form of (defect-ridden) layerwise order
along the flow direction which is inconsistent with the imposed shear
geometry.  In contrast to the preceding case, the system now shows
clear evidence of a finite viscosity at low shear rates (with much
smaller stress fluctuations at low $\gdotbar$). Again, the system
approaches a homogeneous laminar flow at high enough flow rates where
the stress similarly attains the value predicted from the 1D
calculation in this regime.

Finally, for $\zeta= 0.04$, the chaotic state present initially is
perturbed only marginally at small strain rates. The data is
consistent with a finite effective viscosity at low strain rates, but
unlike the previous two cases the flow curve has noticable upward
curvature here (so $\sigma\sim\gdotbar^p$ with $p>1$ cannot be ruled
out). At high strain rates the flow becomes progressively more
homogeneous and the shear stress intersects the 1D curve, signifying a
return to a laminar 1D flow.

\subsubsection{Fixed boundary conditions}

We now turn to the case of fixed boundary conditions. In 1D the results are very similar to the free boundary case \cite{Cates08}. The main difference between fixed and free boundary conditions in 1D is that the superfluid window is somewhat narrowed by the constraint on the order parameter (which inhibits the formation of a band interface close to a wall) and can disappear altogether for large enough value of $l$~\cite{Cates08}. 

As done in the preceding Section focussing on free boundary conditions, we here report flow curves and typical stress snapshots found for different values of activity, which we chose as representative of the steady rolls and the turbulent regimes in the unsheared phase diagram.

\begin{figure}
\begin{center}
\centerline{\includegraphics[width=7.5cm]{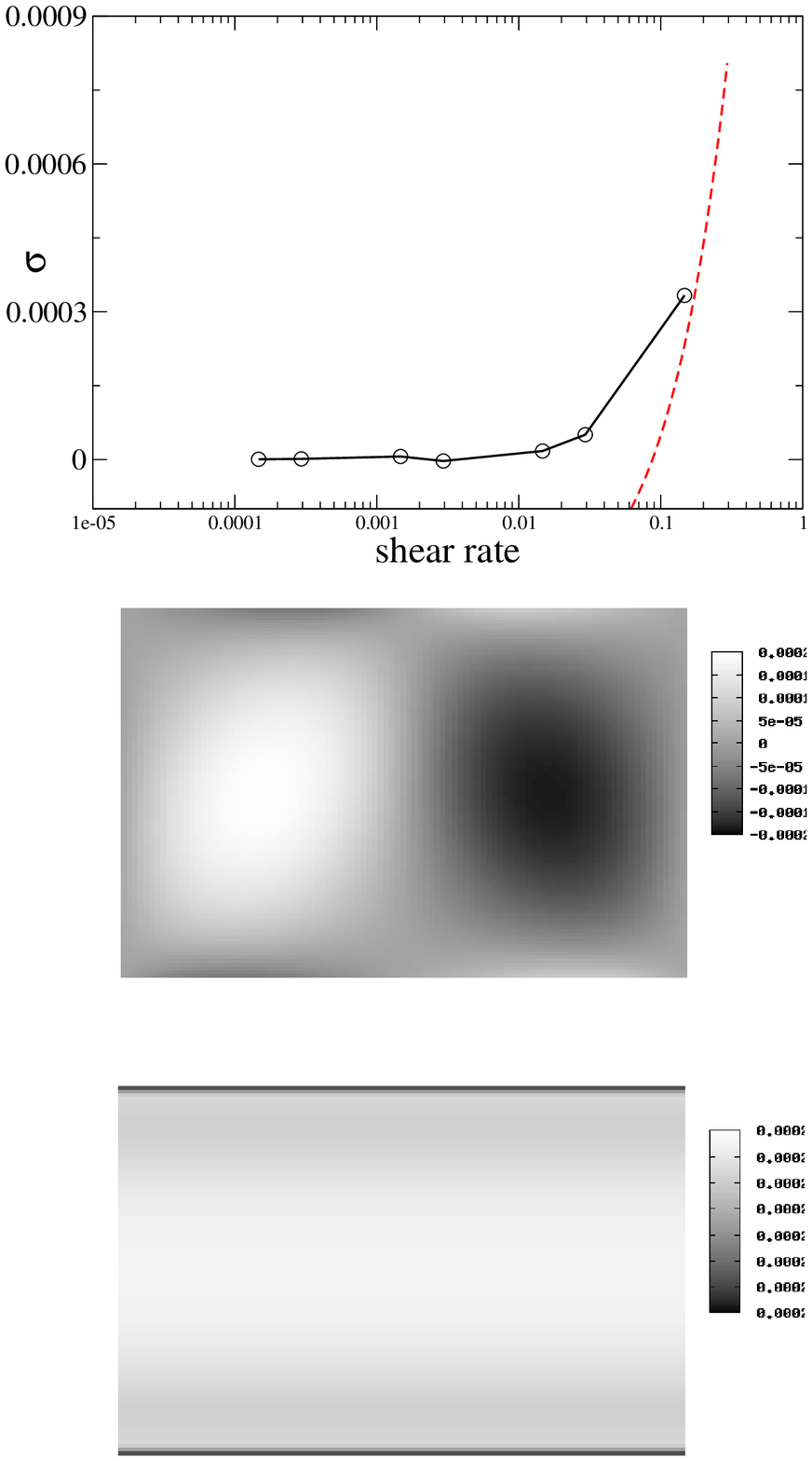}}
\end{center}
\caption{
Top: Flow curve for $\zeta=0.001, l=0.002$.  The dashed curve is
part of the homogeneous constitutive curve.
Lower two panels: representative state snapshot at a late time
for $\dot{\gamma}=10^{-6}$ (second row) and $\dot{\gamma}=10^{-4}$
(third row).}
\label{rolls-fixed}
\end{figure}

\begin{figure}
\begin{center}
\centerline{\includegraphics[width=7.5cm]{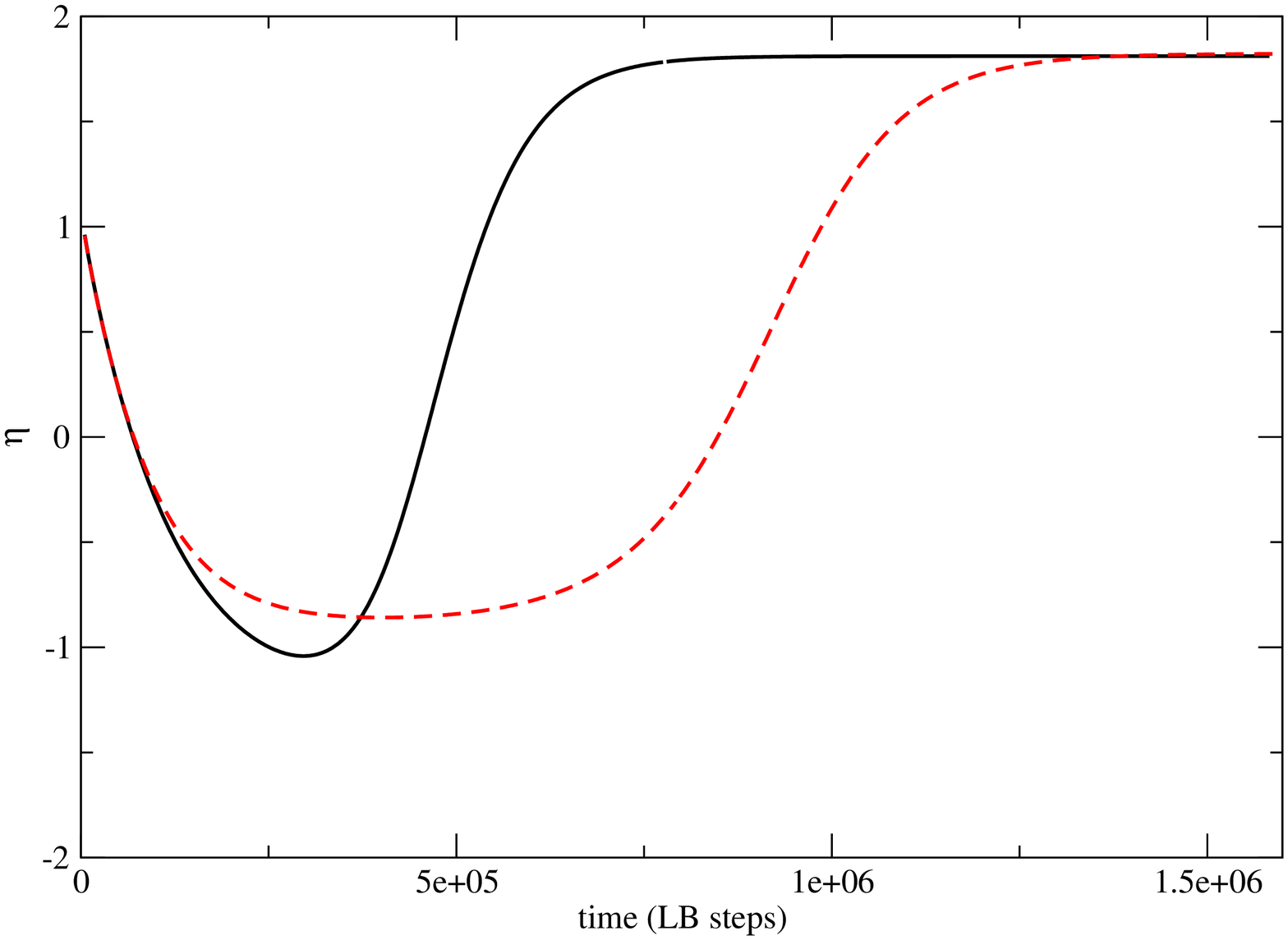}}
\end{center}
\caption{Effective viscosity (normalised by $\eta$) 
versus time for an active gel with
$l=0.002$, $\zeta=0.001$ with
fixed boundary conditions. The solid line refers to
$\dot{\gamma}=0.0002$, the dashed one to $\dot{\gamma}=0.001$.
The steady state effective viscosities are very similar in both cases, proving
that rolls have a linear rheology regime.}
\label{viscosity-rolls-fixed}
\end{figure}

The flow curve observed when shearing the static roll pattern is shown in Fig.~\ref{rolls-fixed}. One important qualitative difference with respect to the free boundary case is that there appears to be a linear regime in this case, in which the viscosity of the rolls is well defined, and larger than $\eta$ -- see Fig.~\ref{viscosity-rolls-fixed}. The extent of this linear regime is very small, and quite likely it shrinks with increasing $L_x$ -- which may explain this qualitative difference between the free and fixed boundary observations. On the other hand, in good agreement with the free boundary case, at high enough shear rates a 1D flow curve is recovered. At low $\dot{\gamma}$, again as with free boundaries, the shear stress takes both positive and negative values. 

\begin{figure}
\begin{center}
\centerline{\includegraphics[width=7.5cm]{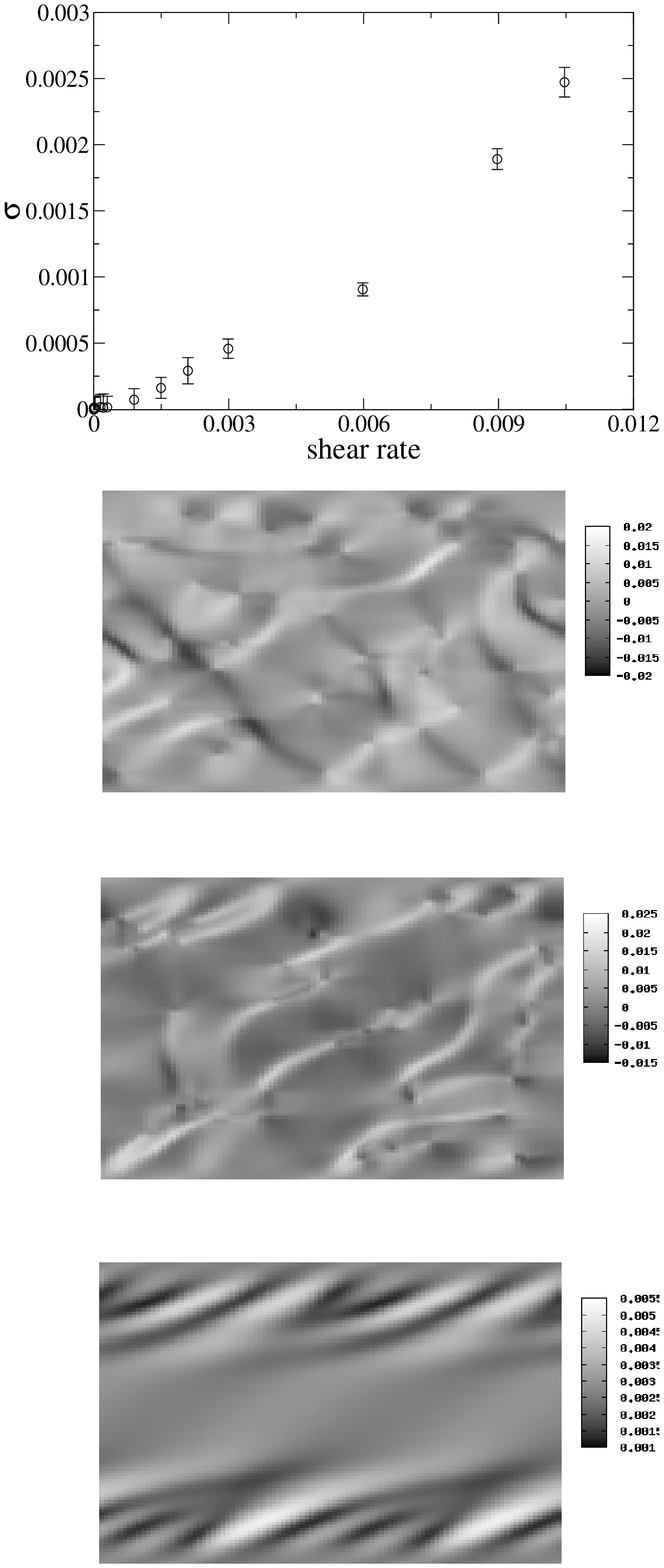}}
\end{center}
\caption{
Top: Flow curve for $\zeta=0.04, l=0.002$.  
Lower three panels: representative snapshots at a late time
for $\dot{\gamma}=0.00005$ (second row), $\dot{\gamma}=0.001$
(third row) and $\dot{\gamma}=0.002$ (bottom row).}
\label{turbulent-fixed}
\end{figure}

The behaviour shown in Fig.~\ref{turbulent-fixed} for $\zeta= 0.04$, is selected as typical for flow curves starting from the unsheared turbulent regime. The flow curve is quantitatively in good agreement with its counterpart for free boundary conditions shown before. The data at small shear rate are consistent with a small yield stress in this case -- this is reasonable as turbulent flows may be expected to dissipate more than rolls, hence have a larger viscosity. The linear regime has also disappeared in the turbulent regime.

\begin{figure}
\begin{center}
\centerline{\includegraphics[width=7.5cm]{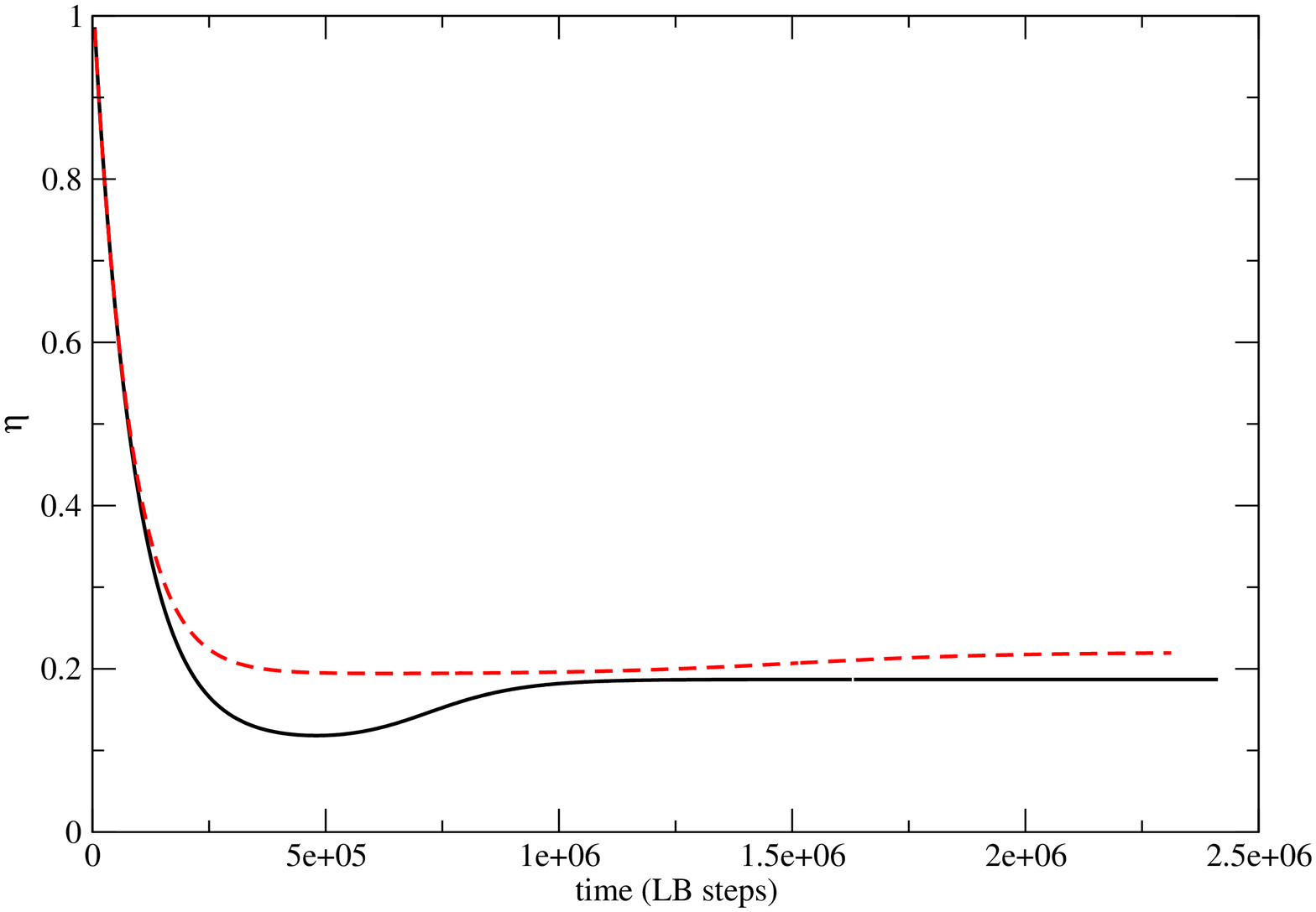}}
\end{center}
\caption{Effective viscosity (normalised by $\eta$) 
versus time for an active gel with
$l=0.002$, $\zeta=0.0005$, 
and fixed boundary conditions. The solid line refers to
$\dot{\gamma}=0.0002$, the dashed one to $\dot{\gamma}=0.001$.
The effective viscosities are smaller than $\eta$ in both cases, and nearly equal (demonstrating the existence of a linear regime).
Note that in order
to get an out-of-plane velocity field in steady state one needs
to start e.g. with an out-of-plane director profile.}\label{out-of-plane-viscosity-fixed}
\end{figure}

\if{
Finally, we recall that the FD simulations employed to obtain the free boundary conditions results are fully 2D, whereas the LB code allows in principle both the director and the velocity field to get an out-of-plane component. This does lead to some differences, although here we mostly focus on the regimes in which the LB leads to in-plane dynamics (this occurs in our case when the initial velocity and director field are both in-plane). However, LB simulations give evidence of a different, out-of-plane, instability, in which the director field rotates out of the plane. This is accompanied by a strong secondary flow. While we do not go into the details of this new state in this work, we note that its flow response is closer to that of a 1D state, and its associated apparent viscosity in the linear regime is this time smaller than $\eta$ (see Fig.~\ref{out-of-plane-viscosity-fixed}). Given the fact that the director goes out of the shear plane in that case, however, one would require 3D simulations to confirm the shape of the flow curves. We leave this to future work.
}
\fi

\section{Conclusion}\label{sec:conclusion}

To conclude, we have presented numerical calculations, by means of both finite difference and Lattice Boltzmann simulations, of the dynamics and flow behaviour of active extensile fluids, such as, for instance, concentrated bacterial suspensions. After checking consistency between the two methods we have used them to explore the different active flows associated with free and fixed boundary conditions. 

In the unsheared case, we found that the transition to spontaneous flow occurs significantly differently than in 1D, consistently with previously reported 2D simulations performed with selected parameters~\cite{SoftMatterReview}. Instead of active bands, we find that the spontaneously flowing phase typically consists of rolls or turbulent flow. In some runs, we were nevertheless able to stabilise tilted active bands, which are however different in shape from the purely 1D {ones.}

The 2D flow curves are also rather different from the 1D ones. Our finite difference simulations with free boundary conditions were performed with a 4:1 aspect ratio between the flow and the flow gradient directions, and this leads to a vanishing or very small linear regime. With fixed boundary conditions, on the other hand, we find that rolls have a well-defined viscosity in the linear regime, and this is larger than the nominal Newtonian viscosity of the fluid. This is in agreement with the expectation that 2D vortices lead to enhanced dissipation with respect to 1D bands. 

{Fully 3D simulations of this problem would undoubtedly be worthwhile, but present serious computational challenges \cite{SoftMatterReview}. An intermediate step to a full 3D simulation is to maintain a two dimensional simulation domain but allow the simulated velocities and director fields to be fully 3D. This is actually what we do in our LB studies (see Appendix) although in all the states discussed so far the flow and director field remains two dimensional, provided that the initial configuration is purely two-dimensional. If this is not the case, a 3D flow may result, where both the director and the velocity field acquire an out-of-plane component (an example is shown in Fig.~\ref{out-of-plane-viscosity-fixed}). Intriguingly this has a linear viscosity which is {\em smaller} than the Newtonian one, in line with the 1D results and dissimilar to the main trends observed in 2D. The details of this out-of-plane flow mode will be pursued elsewhere.} 
 
All our findings can be summed up to say that the dynamics and rheology of active extensile fluids is extremely nonlinear, and that one should take any prediction coming from purely 1D flow, which are typically considered in analytical calculations, with care, as 2D results are often different and more complicated. Nevertheless, our results offer firm predictions for macrorheological experiments on thin 2D samples, which could be performed by e.g. shearing a {\it B. subtilis} monolayer, and it would be very interesting to compare experimental rheological curves to our predictions for the rheology of active rolls and in-plane turbulent states. 

{\bf Acknowledgements} We acknowledge illuminating discussions with A. Morozov. Work funded by EPSRC Grants EP/030173/1 and EP/E05336X/2. MEC holds a Royal Society Research Professorship.

{
\appendix\section{Algorithms}
\label{sec:simulation}
}
In our earlier work on these systems in 1D \cite{Cates08}, the simulations for free and fixed boundary conditions were done using two entirely different algorithms. (This was for historical reasons.) After some initial benchmarking to check there were no systematic discrepancies between algorithms even in 2D, it has proved convenient to continue with this division of labor, although in principle either method should work equally well in both cases.

Accordingly the work reported {in this paper} on free boundary conditions was done with a finite difference (FD) code developed from that previously used in \cite{SMFFD}. That on fixed boundary conditions used instead a hybrid lattice Boltzmann (LB) code, in which FD for the order parameter is coupled to an LB description of momentum transport. This code was previously described in \cite{active_pre}.  

There are additional technical differences between these two sets of simulations. First, many of our LB simulations study square cells, $L_x=1.0$, while our FD simulations
study elongated cells, $L_x=4.0$. Secondly, the FD work assumes a strictly 2D velocity field, so that the out-of-plane component $v_z$ vanishes everywhere. In the LB runs, $v_z(x,y)$ can be nonzero although it vanishes at the walls via the no-slip condition there. This distinction is only relevant when the system spontaneously acquires a flow along $z$. {This may prove relevant in future work as discussed in Section \ref{sec:conclusion}.}

We are interested in flows at low Reynolds number (Re) in which inertial effects are negligible. Within our FD simulations we therefore set $\rho=0$ directly, achieving zero Re from the outset. In the LB approach, finite Re is inevitable since the algorithm exploits the physics of fluid inertia to evolve the dynamics: put differently, strictly zero Re requires strictly zero timestep. 
Therefore, in the units defined above, the fluid density in the LB simulations is
not zero but $\rho=2.3 \times 10^3$. 
Despite this, Re values at the microscopic length scale $l$ remain of order
 $0.1$ or less in our simulations, small enough to maintain the inertialess character of the resulting flow  \cite{CatesCodefReview}. (Even at the box scale, Re generally remains of order 1; moreover comparison with the FD results gives no suggestion that inertia becomes important here either.)


\begin{thebibliography}{99}
\bibitem{Ramaswamy} Y. Hatwalne, S. Ramaswamy, M. Rao, R. A. Simha,
{\it Phys. Rev. Lett.} {\bf 92}, 118101 (2004); R.A. Simha, S. Ramaswamy,
{\it Phys. Rev. Lett.} {\bf 89}, 058101 (2002).
\bibitem{Toner} J. Toner, Y. H. Tu, S. Ramaswamy,
{\it Ann. Phys.} {\bf 318}, 170 (2005).
\bibitem{kruse1} K. Kruse, J. F. Joanny, F. Julicher, J. Prost,
K. Sekimoto, {\it Eur. Phys. J. E} {\bf 16}, 5 (2005); K. Kruse {\it et al.}, {\it Phys. Rev. Lett.}
{\bf 92}, 078101 (2004).
\bibitem{nedelec} F. J. Nedelec {\it et al.}, {\it Nature} {\bf 389},
305 (1997).
\bibitem{Voituriez} R. Voituriez, J. F. Joanny, J. Prost,
{\it Phys. Rev. Lett.} {\bf 96}, 028102 (2006).
\bibitem{Aranson} 
A. Sokolov, I. S. Aranson, J. O. Kessler, R. E. Goldstein,
{\it Phys. Rev. Lett.} {\bf 98}, 158102 (2007).
\bibitem{goldstein} C. Dombrowski, L. Cisneros, S. Chatkaew, R.E. Goldstein, J.O. Kessler, {\it Phys. Rev. Lett.},  {\bf 93}, 098103 
(2004).
\bibitem{Marenduzzo07} D. Marenduzzo, E. Orlandini, J. M. Yeomans,
{\it Phys. Rev. Lett.} {\bf 98}, 118102 (2007).
\bibitem{Cates08} M. E. Cates, S. M. Fielding, D. Marenduzzo, 
E. Orlandini, J. M. Yeomans, {\it Phys.
Rev. Lett.} {\bf 101}, 068102 (2008).
\bibitem{Giomi08} L. Giomi, M. C. Marchetti, T. B. Liverpool, 
{\it Phys. Rev. Lett.} {\bf 101}, 198101 (2008).
\bibitem{Giomi10} L. Giomi, M. C. Marchetti, T. B. Liverpool, 
{\it Phys. Rev. E} {\bf 81}, 051908 (2010).  
\bibitem{Baskaran09} A. Baskaran, M. C. Marchetti, {\it Proc. Natl. Acad. Sci. USA} {\bf 106},
15567 (2009).
\bibitem{EPL} R. Voituriez, J. F. Joanny, J. Prost,
{\it Europhys. Lett.} {\bf 70}, 404 (2005).
\bibitem{Liverpool} T. B. Liverpool, M. C. Marchetti, {\it Europhys. Lett.}
{\bf 69}, 846 (2005).
\bibitem{Liverpool06}   T. B. Liverpool, M. C. Marchetti, 
{\it Phys. Rev. Lett.} {\bf 97}, 268101 (2006).
\bibitem{ramaswamy3} V. Narayan, N. Menon, S. Ramaswamy,
{\it J. Stat. Mech.: Theory and Experiment} P01005 (2006).
\bibitem{llopis} I. Llopis, I. Pagonabarraga, {\it Europhys. Lett.},
{\bf 75}, 999 (2006).
\bibitem{ignacio} S. Ramachandran, P. B. S. Kumar, I. Pagonabarraga,
{\it Eur. Phys. J. E} {\bf 20}, 151 (2006).
\bibitem{SoftMatterReview} M. E. Cates, O. Henrich, D. Marenduzzo, 
K. Stratford, {\it Soft Matter} {\bf 5}, 3791 (2009).
\bibitem{Morozov} A. N. Morozov, D. Marenduzzo, M. E. Cates, in preparation.
{
\bibitem{Wolgemuth} C. Wolgemuth, {\it Biophys. J.} {\bf 95}, 1564 (2008).
\bibitem{Ishikawa} T. Ishikawa and T. J. Pedley, {\it Phys. Rev. Lett.} {\bf 100}, 088103 (2008); T. Ishikawa, J. T. Locsei and T. J. Pedley, {\it J. Fluid Mech.} {\bf 615}, 401 (2008).  
\bibitem{Santillan} D. Saintillan and M. J. Shelley, 
{\it Phys. Rev. Lett.} {\bf 100}, 178103 (2008); {\it Phys. Fluids} {\bf 20}, 123304 (2008).
\bibitem{Graham} P. T. Underhill, J. P. Hernandez-Ortiz and M. D. Graham, {\it Phys. Rev. Lett.} {\bf 100}, 248101 (2008).
\bibitem{Rupert} R. Nash, R. Adhikari, J. Tailleur and M. E. Cates, {\it Phys. Rev. Lett.} {\bf 104}, 258101 (2010).
}
\bibitem{degennes} P.G. de Gennes and J. Prost, {\it The Physics of
Liquid Crystals, 2nd Ed.}, Clarendon Press, Oxford, (1993).
\bibitem{beris}
A.N. Beris and B.J. Edwards, {\it Thermodynamics of Flowing Systems},
Oxford University Press, Oxford, (1994); A.N. Beris, B.J. Edwards and
M. Grmela, {\it J. Non-Newtonian Fluid Mechanics}, {\bf 35} 51 (1990). 
\bibitem{O92}
P.D. Olmsted and P.M. Goldbart, Phys. Rev. A {\bf 46}, 4966 (1992).
\bibitem{O99}
P.D. Olmsted and C.-Y. David Lu, Phys. Rev. E {\bf 56}, 55 (1997);
ibid, {\bf 60}, 4397 (1999).
\bibitem{SMFFD} S. M. Fielding, {\it Phys. Rev. E}
{\bf 77}, 021504 (2008).

\bibitem{active_pre} D. Marenduzzo, E. Orlandini,
M. E. Cates, J. M. Yeomans, {\it Phys. Rev. E} {\bf 76},
031921 (2007).
\bibitem{CatesCodefReview} M. E. Cates {\it et al.}, {\it J. Phys. Cond. Mat.} {\bf 16}, S3903--S3907 (2004).

\if{

\bibitem{bray}  D. Bray, {\em Cell Movements: from Molecules to
Motility}, Garland Publishing, New York (2000).
\bibitem{peter} P. R. Cook, {\it Principles of Nuclear Structure and
Function}, Wiley, New York (2001).
\bibitem{beads} J. van der Gucht, E. Paluch, C. Sykes,
{\it Proc. Natl. Acad. Sci. USA} {\bf 102}, 7847 (2005);
M. F. Carlier {\it et al.}, {\it Bioessays} {\bf 25}, 336 (2003).
\bibitem{cytoskeleton} J. Howard, {\it Mechanics of Motor Proteins
and the Cytoskeleton}, Sinauer Associates, Inc., Sunderland (2001).
\bibitem{active_actin_myosin} D. Humphrey {\it et al.},
{\em Nature} {\bf 416}, 413 (2002).
\bibitem{surrey} T. Surrey, F. Nedelec, S. Leibler and E. Karsenti, 
{\it Science} {\bf 292}, 1167 (2001).
\bibitem{lubensky}  C. Storm, J.J. Pastore, F.C. MacKintosh, 
T.C. Lubensky, P.A. Janmey, {\it Nature} {\bf 435}, 191 (2005).
\bibitem{activeLB} D. Marenduzzo, E. Orlandini, J. M. Yeomans, 
{\it Phys. Rev. Lett.} {\bf 98}, 118102 (2007).
D. Marenduzzo, E. Orlandini, M. E. Cates,
J. M. Yeomans, {\it J. Non-Newt. Fluid Mech.} {\bf 149}, 56 (2008).
\bibitem{forest} M. G. Forest, R. H. Zhou, Q. Wang,
{\it Phys. Rev. Lett.} {\bf 93}, 088301 (2004).
\bibitem{ramaswamy_chaos}  B. Chakrabarti, M. Das, C. Dasgupta, 
S. Ramaswamy, A. K. Sood, {\it Phys. Rev. Lett.} {\bf 92}, 055501
(2004).
\bibitem{mike} A. Aradian, M.E. Cates, {\it Europhys. Lett.}
{\bf 70}, 397 (2005).
\bibitem{chlamydomonas} M. Polin {\it et al.}, {\it Science} {\bf 325},
487 (2009).
\bibitem{thoumine} O. Thoumine, A. Ott, {\it J. Cell. Sci.}
{\bf 110}, 2109 (1997).
\bibitem{frey} J. Uhde, M. Keller, E. Sackmann, A. Parmegiani,
E. Frey, {\it Phys. Rev. Lett.} {\bf 93}, 268101 (2004).
\bibitem{succi} S. Succi, {\it The Lattice Boltzmann Equation}, 
Oxford University Press (2001).
\bibitem{colin} C. Denniston, E. Orlandini, J. M. Yeomans,
{\it Phys. Rev. E} {\bf 63}, 056702 (2001).
\bibitem{lblc} C. Denniston, D. Marenduzzo, E. Orlandini, 
J. M. Yeomans, {\it Phil. Trans. R. Soc. Lond. A} {\bf 362}, 1745 (2004).
\bibitem{nidhal} N. Sulaiman, D. Marenduzzo, J.M. Yeomans, 
{\it Phys. Rev. E} {\bf 74}, 041708 (2006).
\bibitem{M82}
G. Marrucci, Mol. Cryst. Liq. Cryst. {\bf 72}, 153 (1982).
\bibitem{LandauLifshitz} L.D. Landau and E.M. Lifshitz, {\it Theory of
Elasticity, 3rd Ed.}, Pergamon Press, Oxford, (1986).
\bibitem{Zapo95} M. Zapotocky, P.M. Goldbart and N. Goldenfeld,
Phys. Rev. E {\bf 51}, 1216 (1995).
\bibitem{rheoHAN} D. Marenduzzo, E. Orlandini, J. M. Yeomans,
{\it Europhys. Lett.} {\bf 64}, 406 (2003).
}\fi

\end{thebibliography}
\end{document}